\documentclass[11pt,a4paper]{article}           
\usepackage{amsmath, amsfonts,amssymb,mathtools,nicefrac,nccmath,cases}
\usepackage{microtype,booktabs,multirow}
\usepackage[usenames,dvipsnames,table]{xcolor} 
\usepackage{cite}
\usepackage{colortbl}
\usepackage{tikz}
\usetikzlibrary{}
\usetikzlibrary{shapes,shapes.multipart,shapes.callouts,shapes.arrows,shapes.geometric}
\usetikzlibrary{arrows,arrows.meta}
\usetikzlibrary{decorations.pathmorphing,decorations.markings}
\usetikzlibrary{calc,bending,intersections}

\usepackage[linktocpage=true]{hyperref}
\hypersetup{colorlinks=true,linkcolor=nicecolor,citecolor=nicecolor,urlcolor=nicecolor}
\definecolor{nicecolor}{rgb}{0.1, 0.3, 0.4}

\definecolor{blue}{rgb}{0.06, 0.3, 0.57}
\definecolor{Gray}{gray}{0.4}

\definecolor{nicecolor}{rgb}{0.1, 0.3, 0.4}
\definecolor{blue}{rgb}{0.06, 0.3, 0.57}
\definecolor{Gray}{gray}{0.4}
\colorlet{tableheadcolor}{gray!15} 
\newcommand{\headcol}{\rowcolor{tableheadcolor}} 
\colorlet{tablerowcolor}{gray!7} 
\newcommand{\topline}{\arrayrulecolor{black}\specialrule{0.1em}{\abovetopsep}{0.5pt}%
	\arrayrulecolor{tableheadcolor}\specialrule{\belowrulesep}{0pt}{-3pt}%
	\arrayrulecolor{black}}
\newcommand{\midline}{\arrayrulecolor{tableheadcolor}\specialrule{\aboverulesep}{-1pt}{0pt}%
	\arrayrulecolor{black}\specialrule{\lightrulewidth}{0pt}{0pt}%
	\arrayrulecolor{white}\specialrule{\belowrulesep}{0pt}{-3pt}%
	\arrayrulecolor{black}}
\newcommand{\bottomline}{\arrayrulecolor{white}\specialrule{\aboverulesep}{0pt}{-2pt}%
	\arrayrulecolor{black}\specialrule{\heavyrulewidth}{0pt}{\belowbottomsep}}%

\def\hybrid{\topmargin -20pt    \oddsidemargin 0pt
	\headheight 0pt \headsep 0pt
	\textwidth 6.5in        
	\textheight 9in         
	\textwidth 6.25in       
	\textheight 9 in       
	\marginparwidth .875in
	\parskip 5pt plus 1pt 
	\jot = 1.5ex
}
\hybrid
\numberwithin{equation}{section}
\numberwithin{table}{section}
\newcommand{\beq}{\begin{equation}}  \newcommand{\eeq}{\end{equation}}
\newcommand{\bal}{\begin{aligned}}   \newcommand{\eal}{\end{aligned}}
\newcommand{\bea}{\begin{eqnarray}}  \newcommand{\eea}{\end{eqnarray}}
\def\beqa{\begin{eqnarray}}
\def\eeqa{\end{eqnarray}}

\newcommand{\bmat}{\left(\begin{array}}
	\newcommand{\emat}{\end{array}\right)}

\newcommand{\pairing}{\vartheta}
\newcommand{\base}{\eta}

\newcommand{\K}[1]{\c K_{#1}}
\DeclareMathOperator{\rk}{rk}
\DeclareMathOperator{\ch}{ch}
\newcommand{\ve}{\varepsilon}

\newcommand{\np}{{(n)}}

\newcommand{\ttext}[1]{\ \ \text {#1} \ \ }
\newcommand{\qtext}[1]{\quad \text{#1} \quad}
\newcommand{\qqtext}[1]{\qquad \text{#1} \qquad}
\newcommand{\Qtext}[1]{\qquad \text{#1} \quad}

\newcommand{\cO}{\mathcal{O}}

\newcommand{\cC}{\mathcal{C}}

\newcommand{\cK}{\mathcal{K}}
\newcommand{\cN}{\mathcal{N}}

\newcommand{\cI}{\mathcal{I}}
\newcommand{\cJ}{\mathcal{J}}
\newcommand{\cR}{\mathcal{R}}

\newcommand{\cV}{\mathcal{V}}

\newcommand{\cM}{\mathcal{M}}

\newcommand{\I}{\text{Im}}

\newcommand{\be}{\begin{equation}}
\newcommand{\ee}{\end{equation}}

\newcommand{\rII}{\mathrm{II}}
\newcommand{\rIII}{\mathrm{III}}
\newcommand{\rIV}{\mathrm{IV}}

\DeclareMathOperator{\dd}{d \!}
\renewcommand{\d}{\dd \, \!}

\newcommand{\oneloop}{\text{1-loop}}

\newcommand{\Z}{\mathbb Z}

\newcommand{\ws}{\wedge \star}
\newcommand{\wsh}{\wedge \hat \star}

\newcommand{\vabs}[1]{\left | #1 \right |}

\renewcommand{\c}{\mathcal }

\renewcommand{\b}{\beta}
\renewcommand{\a}{\alpha}

\newcommand{\tr}[1]{{\rm Tr} \left( #1\right)}

\newcommand{\abs}[1]{\left | #1 \right |}

\newcommand{\mmfrac}[2]{\mfrac{\raisebox{-2pt}{$#1$}}{#2}}

\newcommand{\bb}{{\rm b}}
\renewcommand{\o}{\omega}

\renewcommand{\L}{\Lambda}
\newcommand{\D}{\Delta}

\newcommand{\mpl}[1]{M_{\rm pl,#1}}
\renewcommand{\mp}{M_{\rm pl}}
\newcommand{\nvf}{n_V^{(5)}}

\newcommand{\qo}{\mathbf{q}_0}
\newcommand{\hoo}[1]{h^{1,1}(#1)}
\newcommand{\N}[1]{N_{(#1)}}
\newcommand{\two}{{(2)}}
\newcommand{\T}{{\rm T}}
\newcommand{\type}[1]{{\sf Type\ A}_{(#1)}}

\newcommand{\KK}[2]{\mathcal K_{#1}^{(#2)}}
\newcommand{\norm}[1]{\left\lVert#1\right\rVert}
\renewcommand{\c}{\mathcal }

\renewcommand{\b}{\beta}
\renewcommand{\a}{\alpha}
\newcommand{\w}{\mathbf w}
\renewcommand{\v}{\mathbf v}
\renewcommand{\u}{\mathbf u}
\newcommand{\x}{\mathbf x}
\newcommand{\q}{\mathbf q}
\newcommand{\KKbis}[2]{\c K _{#1}^{[#2]}}

\begin{document}
	
	\baselineskip=14pt
	\parskip 5pt plus 1pt

	\vspace*{-1.5cm}
	\begin{flushright}    
		{\small 
			
		}
	\end{flushright}
	
	\vspace{2cm}
	\begin{center}        
		{\LARGE The Swampland Distance Conjecture for K\"ahler moduli}
	\end{center}
	
	\vspace{0.5cm}
	\begin{center}        
		{\large  Pierre Corvilain$^{1}$, Thomas W.~Grimm$^{1}$, Irene Valenzuela$^2$}
	\end{center}
	
	\vspace{0.15cm}
	\begin{center}        
		\emph{$^1$ Institute for Theoretical Physics \\
			Utrecht University, Princetonplein 5, 3584 CE Utrecht, The Netherlands}\\[0.15cm]
		\emph{$^2$ Department of Physics, Cornell University, Ithaca, New York, USA}
	\end{center}
	
	\vspace{2cm}
	
	
	\begin{abstract}
		\noindent
		The Swampland Distance Conjecture suggests that an infinite tower of modes becomes exponentially 
light when approaching a point that is at infinite proper distance in field space. 
In this paper we investigate this conjecture in the K\"ahler moduli spaces of Calabi-Yau threefold compactifications
and further elucidate the proposal that the infinite tower of states is generated by the discrete symmetries 
associated to infinite distance points. In the large volume regime the infinite tower of states is generated 
by the action of the local monodromy matrices and encoded by an orbit of D-brane charges.
We express these monodromy matrices in terms of the triple intersection numbers
to classify the infinite distance points and construct the associated infinite charge orbits that become massless. 
We then turn to a detailed study of charge orbits in elliptically fibered Calabi-Yau threefolds. We argue that for these geometries
the modular symmetry in the moduli space can be used to transfer the large volume orbits to the small elliptic fiber regime.
The resulting orbits can be used in compactifications of M-theory that are dual to 
F-theory compactifications including an additional circle. In particular, we 
show that there are always charge orbits satisfying the distance conjecture that correspond to Kaluza-Klein towers along that circle. 
Integrating out the KK towers yields an infinite distance in the moduli space thereby supporting the idea of emergence in that context.
\end{abstract}

\thispagestyle{empty}
\clearpage

\setcounter{page}{1}


\newpage

\tableofcontents


\section{Introduction}

The term Swampland~\cite{Vafa2005} refers to those quantum  effective field theories which cannot be UV embedded in a consistent theory of quantum gravity. In particular, there are several proposals for consistency constraints that any effective theory weakly coupled to Einstein gravity must satisfy to arise from string theory. In this paper, we will focus on the Swampland Distance Conjecture (SDC)~\cite{Ooguri2007}, for which infinite distances in field space imply an infinite tower of states becoming massless exponentially fast in the proper field distance. This infinite tower of states is associated to a quantum gravity cut-off that goes to zero at infinite distance and above which a quantum effective field theory description weakly coupled to Einstein gravity is no longer possible. Therefore, the conjecture implies an upper bound on the scalar field range that any effective theory can accommodate in terms of the energy scale up to which the effective theory is valid. Such a bound can have 
several potential implications for phenomenology especially when constructing models of large field inflation.

Due to the importance of the swampland criteria to yield non-trivial quantum gravity constraints at low energies as well as to provide new guidelines to make progress in high energy physics, it is essential to gather more evidence to prove or disprove these conjectures in a rigorous way. It is the aim of this paper to continue testing the Swampland Distance Conjecture in string theory compactifications.  As a byproduct of analyzing this conjecture we further elucidate the very rich underlying geometric structure of the moduli space and compactification manifolds required for the conjecture to hold. This structure, together with the understanding of the states arising in string theory, implies highly non-trivial correlations between the number of light states and field distances.  In certain cases, as we will see, the SDC seems to be satisfied in a conspiratorial 
way by string theory. This invites us to continue exploring the SDC to reveal the underlying quantum gravity principle responsible for the validity of the conjecture, and hopefully learn new lessons about quantum gravity itself.

Non-trivial evidence for the SDC was obtained in~\cite{Grimm:2018ohb,Grimm:2018cpv} by studying infinite distance singularities of the complex structure moduli space of Calabi-Yau manifolds. We also refer the reader to~\cite{Lee2018,Lee2019} for a recent general analysis of weak coupling points in F-theory and~\cite{Palti:2015xra,Baume:2016psm,Valenzuela:2016yny,Bielleman:2016olv,Blumenhagen:2017cxt,Palti:2017elp,Hebecker:2017lxm,Cicoli:2018tcq,Blumenhagen:2018nts,Buratti:2018xjt}  for a series of previous works analyzing the conjecture in concrete string compactification setups. The power of the approach followed in~\cite{Grimm:2018ohb,Grimm:2018cpv} was its model independence as the results are valid regardless the specific Calabi-Yau under consideration. The proposal is to identify the infinite tower of states with an infinite charge orbit generated by a monodromy action of infinite order. This infinite order monodromy is a necessary condition for a singular locus to be at infinite field distance. 
In one-parameter degenerations the charge orbit was shown in~\cite{Grimm:2018ohb} to be populated by an exponentially increasing number of BPS states that become exponentially light as we approach the singular locus, providing evidence for the conjecture. Furthermore, it was proposed that the infinite field distance itself emerges from quantum corrections of integrating out the infinite tower of states. We will revisit this argument and provide a general field theory computation that highlights the properties that need to be met by the tower of states. Furthermore, the monodromy transformation is translated to an axionic discrete shift symmetry in the effective theory which enhances to a continuous shift symmetry at infinite distance. This provides a new understanding of the SDC as a quantum gravity obstruction to restore global symmetries. Everything fits together in a beautiful story linked to the monodromy action. The next obvious question is how much of this story can be generalized to other moduli spaces.

In this paper, we will explore the Swampland Distance Conjecture in the multi-dimensional K\"ahler moduli space of Calabi-Yau compactifications. We will show how the techniques introduced in~\cite{Grimm:2018ohb,Grimm:2018cpv} can also be used to identify an infinite charge orbit becoming massless at infinite distance in K\"ahler moduli spaces. The main focus of our paper will be on the study of infinite distance loci and charge orbits at the large volume regime. The monodromy action can be written in full generality in terms of the intersection numbers and topological data of the Calabi-Yau threefold, allowing for a classification of the infinite distance limits at large volume. We will also provide the general result for the infinite charge orbit becoming massless at these limits. The existence of such orbits was shown in~\cite{Grimm:2018cpv}, where it was also argued that this crucially requires to address the issue of path-dependence by
applying the powerful mathematical machinery 
of~\cite{CKS}. However, in this work we will be able to determine the charge orbit by studying a comparably simple set of vector equations. This refined approach is valid more generally and can also be applied to the complex structure moduli space. 
Subsequently we will discuss the interesting phenomenon of transferring the charge orbit to other infinite distance points of the moduli space away from large volume. In the case of elliptic fibrations, it is possible to carry the charge orbit from large volume to the small fiber point by applying double T-duality along the fiber.

The tower of states becomes exponentially light with respect to the Planck scale. This means that, if we are moving along some path in the moduli space which is also sending~$\mp \rightarrow \infty$, they can become very heavy while still satisfying~$m/\mp\rightarrow 0$. This is a result of the fact that the SDC only gives non-trivial implications in the IR effective theory if the~$\mp$ is forced to remain finite, while all implications go away when gravity decouples. This feature is particularly visible when moving in the K\"ahler moduli space, since~$\mp\rightarrow \infty$ at large volume. Furthermore, there can also be more than one tower of states becoming exponentially light with respect to~$\mp$ at infinite distance. For instance, if we consider type IIA compactified on a Calabi-Yau threefold, we get that the infinite charge orbits generated by the monodromy action at large volume consist of a tower of particles arising from bound states of D0-D2 branes. Clearly, there will also be Kaluza-Klein towers of states becoming massless at large volume. However, it is the tower of D0-D2 branes that appears to be relevant for the proposals of emergence and global symmetries in the K\"ahler moduli space. In particular, the infinite field distance can be understood as emerging from quantum corrections of integrating out D-brane states rather than Kaluza-Klein states in this case. Notice also that if the infinite distance emerges from integrating out the tower of states, this emergence interpretation should be equally applicable for the intersection numbers and topological discrete data of the Calabi-Yau manifold.

There are other instances, though, in which a Kaluza-Klein tower can be responsible for (at least part of) the infinite field distance. This is, for example, the case in the circle compactification performed in order to implement the duality of M-theory and F-theory. The 6D effective theory of F-theory compactified on an elliptically fibered Calabi-Yau threefold can be derived from compactifying M-theory on the same Calabi-Yau manifold to five dimensions and sending the volume of the fiber to zero. The limit of shrinking the elliptic fiber corresponds to decompactifying an additional circle and opening up an extra dimension in the F-theory side. It is known that quantum corrections from the KK tower in the circle F-theory compactification are essential to match with the classical M-theory reduction~\cite{Bonetti:2011mw,Bonetti:2012fn,Bonetti:2013ela}.\footnote{It was recently shown in~\cite{Grimm:2018weo} that this infinite tower of states is also crucial in order to account for the entropy of certain F-theory black holes.} In this paper, we will also analyze the infinite distance limits in the M-theory geometry, and recover the KK tower of the circle compactification of F-theory from following the infinite charge orbit to the small fiber regime in M-theory. This provides a geometric realization of the Kaluza-Klein tower in terms of a charge orbit generated by a monodromy action of infinite order.

The outline of the paper goes as follows. We will start in section~\ref{swampland_gs} discussing the general properties that the tower of states must satisfy and revisiting the idea of emergence. We also present a new field theory computation that shows how quantum corrections from integrating out any infinite tower up to its species scale generates an infinite field distance (and consequently, an exponential mass behavior) as long as the number of species increases as we move in field space. We will then discuss the microscopic meaning of the species bound in Kaluza-Klein compactifications. In section~\ref{TypeIIAorbits} we will analyze infinite distances in the large volume regime of Calabi-Yau threefold compactifications. We will construct the infinite charge orbits becoming massless at the different large volume limits and their microscopic interpretation in terms of type IIA string theory. We will also discuss how to carry the charge orbit to the small fiber volume in elliptic fibrations. In section~\ref{MF-section} we will discuss infinite distances and charge orbits arising in the duality between M-theory and F-theory, providing a geometric realization for the KK tower in terms of an infinite charge orbit in M-theory. Finally, section~\ref{conclusions} contains our conclusions.

\section{Swampland, emergence of infinite distance and global symmetries} \label{swampland_gs}

Consider the moduli space of a consistent quantum gravity effective theory parametrized by the expectation values of the massless scalar fields in the theory. The Swampland Distance Conjecture~\cite{Ooguri2007} states that any low energy effective theory defined at a particular point of the moduli space is only valid in a finite domain around that point, because there will be an infinite tower of states becoming exponentially light when moving infinitely far away and signaling the complete breakdown of the effective theory. More concretely, when starting with an effective theory defined at a point Q in the moduli space and moving towards another point $P$, the mass of this tower of states behaves as
\begin{equation}
m(P)\sim m(Q)e^{-\gamma \, d(P,Q)}
\label{SDC}
\end{equation}
in the limit~$d(P,Q)\rightarrow \infty$.  Here,~$d(P,Q)$ is the geodesic distance between the two points, and~$\gamma$ is a positive constant which is not specified in generality. This infinite tower implies the complete breakdown of the effective theory in the sense that quantum gravitational effects become important and a quantum field theory description with infinitely many fields weakly coupled to Einstein gravity is not possible. 
Therefore, not only the low energy effective theory breaks down because of the presence of new states, but the quantum gravity cut-off~$\Lambda_{QG}$ also goes to zero exponentially fast.
As it stands, the conjecture leaves many open questions: Can we universally specify~$\gamma$ and~$\Lambda_{QG}$? How do they change if we move along a non-geodesic trajectory? Is there any universal prescription to identify the tower of states? What is the underlying quantum gravity principle which forces the conjecture to hold?
For the latter question, there are two recent proposals:
\begin{itemize}
	\item The infinite distance itself emerges from quantum corrections of integrating out the infinite tower of states up to the species bound of the tower~\cite{Ooguri2007,Grimm:2018ohb,Heidenreich2018a}.
	\item The infinite tower is a quantum gravity obstruction to restore a global symmetry at the infinite distance limit~\cite{Grimm:2018ohb}.
\end{itemize}
These two proposals find confirmation~\cite{Grimm:2018ohb} at the infinite distance loci of the complex structure moduli space of Type IIB Calabi-Yau compactifications, where it was also proposed a general prescription to identify the tower of states in terms of a charge orbit generated by a monodromy action of infinite order. It is the aim of this paper to extend the discussion to K\"ahler moduli spaces, and to check whether these two proposals, as well as the aforementioned prescription to identify the tower, are still valid. Before turning to do so, we will first explain in more detail and revisit these two proposals in view of the new insights gathered in this paper. Let us remark, though, that the following discussion in this section is empty without the solid technical work that follows in section~\ref{TypeIIAorbits} and~\ref{MF-section}. Furthermore, since moving in the K\"ahler moduli space usually also implies varying the Planck mass, there are some subtleties that need to be addressed. Hence, we will first discuss these subtleties in section~\ref{sec:KK} in a toy model example: a circle Kaluza-Klein compactification.

\subsection{Emergence and global symmetries}
In the following we will describe in more detail the above two proposals and present a new computation that shows how the exponential mass behavior (and the infinite field distance) is an automatic consequence of integrating out  any infinite tower of states (regardless their concrete mass) up to the species bound of the tower, as long as the tower gets compressed as we move in the moduli space. This leads to a natural identification of the quantum gravity cut-off with the species bound, as we will next discuss.

\paragraph{Emergence of infinite distance from integrating out a tower}
Let us consider a~$D$-dimensional effective theory of a massless scalar field~$\phi$ plus a tower of heavy particles~$h$ whose mass depends on~$\phi$ as~$m_n(\phi)=n\Delta m(\phi)$. 
We will follow very closely~\cite{Grimm:2018ohb,Heidenreich2018a} but without assuming any particular form for~$\Delta m(\phi)$. The power of our results will precisely reside in this independence of the form of~$\Delta m(\phi)$. 
The Lagrangian is
\begin{equation}
\label{effthemulh}
\mathcal{L}= \mmfrac 12 \left(\partial \phi \right)^2 + \sum_n \left[  \mmfrac 12 \left(\partial h_n \right)^2 + \mmfrac 12 m_n\left(\phi\right)^2h_n^2 \right]\;.
\end{equation}
We are interested in the quantum corrections to the field metric of~$\phi$ when integrating out the massive infinite tower of states. 
However, any tower of states weakly coupled to Einstein gravity has an associated cut-off scale above which quantum gravitational effects become important and the quantum field theory description breaks down. Since the procedure of integrating out can only be performed within the realm of an effective quantum field theory, we should only integrate out the states up to 
this quantum gravity scale~$\L_{QG}$. There is a very natural candidate for~$\Lambda_{QG}$ known as the species scale~\cite{ArkaniHamed:2005yv,Distler:2005hi,Dimopoulos:2005ac,Dvali:2007wp,Dvali:2007hz},
\begin{align} \label{SpeciesBound}
\L_{QG}\simeq \frac{\mpl{D}}{N_{QG}^{\frac{1}{D-2}}},
\end{align}
where~$N_{QG}$ is the number of species (i.e. elementary particles weakly coupled to gravity) present below the energy scale~$\L_{QG}$, and~$\mpl{D}$ is the~$D$-dimensional Planck mass. For the above tower of particles of evenly increasing mass, we have
\begin{equation}
N_{QG}=\frac{\L_{QG}}{\Delta m(\phi)}
\end{equation}
implying
\begin{equation}
\L_{QG}\simeq \left (\mpl{D}^{D-2}\Delta m(\phi) \right )^{\frac{1}{D-1}}\qqtext{and} N_{QG} = \left(\frac{\mpl D}{ \Delta m(\phi)}\right)^{\frac{D-2}{D-1}} \ .
\end{equation}
Therefore, if~$\Delta m$ depends on the point of the moduli space parametrized by~$\phi$, so will the species scale. In fact if we now consider that the whole tower becomes massless at a particular point~$\phi_0$, so~$\Delta m_n(\phi_0) =0$ and~$N_{QG}\rightarrow \infty$, the species scale will go to zero at that point, i.e.~$\L_{QG}(\phi_0)= 0$.

We can now compute the one-loop quantum corrections to the field metric of~$\phi$ when integrating out the tower of massive states, given by~\cite{Ooguri2007,Grimm:2018ohb,Heidenreich2018a}
\begin{equation}
\label{goneloop}
g_{\phi\phi}^{\oneloop} 
\sim \sum_n\, m_n(\phi)^{D-4} \, (\partial_\phi m_n(\phi)) ^2 
\end{equation}
When summing only over the number of species below~$\Lambda_{QG}$, we get
\begin{align}
g_{\phi\phi}^{\oneloop} 
\sim N_{QG}^{D-1} \, \Delta m(\phi)^{D-4} \, \big (\partial_\phi \Delta m(\phi) \big ) ^2\sim \mpl{D}^{D-2} \,\left( \frac{\partial_\phi \Delta m (\phi)}{\Delta m(\phi)} \right)^2. \label{g1Lgen}
\end{align}
The distance between two points of the moduli space~$\phi_0$ and~$\phi_1$ is then given by 
\begin{align}
d(\phi_0,\phi_1) = \int_{\phi_0}^{\phi_1} \sqrt{g_{\phi\phi}} \sim \log \left( \frac{\Delta m(\phi_1)}{\Delta m(\phi_0)}\right)
\end{align}
which indeed diverges if~$\Delta m(\phi_1) \to 0$, and the masses decrease exponentially as we approach the infinite distance point,
\begin{align}
\Delta m(\phi_0) \sim \Delta m(\phi_1) \, e^{-\gamma \, d(\phi_0,\phi_1)}.
\end{align}
where~$\gamma$ encodes all the numerical factors that we have neglected in the above procedure of integrating out and that will depend on the properties of the tower. 
Notice that we did not need to specify the dependence of the masses on~$\phi$. The logarithmic divergence of the proper field distance, and consequently the exponential mass behavior, emerges from integrating out any tower of states up to its species bound\footnote{See \cite{Harlow:2015lma,Heidenreich2018a,Grimm:2018ohb,Harlow2018a} for the proposal that the Weak Gravity Conjecture is also implied by the idea that the small gauge coupling emerges from integrating out the massive charged WGC states up to the species bound.}. The only thing that matters is that the tower gets compressed, i.e.~$\Delta m(\phi_0)$ goes to zero at the point in question.
In terms of the quantum corrected proper field distance, the number of species then increases exponentially and the quantum gravity cut-off decreases exponentially fast,
\begin{equation}
\Lambda_{QG}\sim \mpl{D} \, e^{- \lambda \, d(\phi_0,\phi_1)}
\end{equation}
where~$\lambda\sim \gamma/(D-1)$.
This toy model computation removes part of the mysticism of the conjecture relating infinite distances and infinite towers of states. If the number of species increases when approaching a point of the moduli space, quantum corrections from this tower will automatically generate a logarithmic field distance divergence in terms of the mass of these states. In~\cite{Ooguri2007} it was pointed out that not every infinite massless tower necessarily generates an infinite field distance. We however think that this will always be the case as long as they count as different species.

Finally, there are also two possible levels of emergence. It could either be that the infinite tower generates part of the infinite field distance, a classical divergence being also present, or that the infinite field distance fully emerges from quantum corrections form integrating out the tower. In the latter case, the fact that moduli spaces are in general non-compact would be an IR effect from integrating out infinite towers of states that become massless at particular points. Why these towers should exist will be the question of the next section about global symmetries. As a final remark, notice that quantum corrections will dominate over the classical piece  if the tower of states  satisfies what was called the Scalar Weak Gravity Conjecture~\cite{Palti:2017elp},
\begin{equation}
g_{\phi\phi}^{\oneloop}\geq g_{\phi\phi} \qtext{if} g^{\phi\phi}\bigg( \frac{\partial_\phi \Delta m (\phi)}{\Delta m(\phi)} \bigg)^2\gtrsim 1
\end{equation}
where we have used \eqref{g1Lgen}.
Equivalently, the Scalar WGC is automatically satisfied if the idea of emergence holds. This also provides a motivation to have~$\gamma,\lambda\gtrsim 1$.

\paragraph{Obstruction to global symmetries}
A nice relation between the SDC and the absence of global symmetries was proposed in~\cite{Grimm:2018ohb}. As we will explain later on in more detail, the infinite tower of states is identified with a charge orbit generated by a discrete monodromy transformation of infinite order. When reaching the infinite distance point, this discrete transformation enhances to a continuous one, which would imply the presence of a continuous global shift symmetry in the effective theory. The presence of the infinite tower, which automatically forces the quantum gravity cut-off to go to zero, can then be understood as a quantum gravity obstruction to restore this global symmetry. This is consistent with the common lore that global symmetries are not allowed in quantum gravity (recently proved in the context of AdS/CFT~\cite{Harlow2018,Harlow2018a}). The key point is that the conjecture states how the effective theory breaks down when trying to recover a global symmetry in a continuous way. Therefore, it quantifies how approximate a global shift symmetry can be, by providing a quantum gravity cut-off above which no effective field theory enjoying that approximate global symmetry is valid. It also nicely connects with the Weak Gravity Conjecture \cite{ArkaniHamed:2006dz}, which analogously quantifies what goes wrong when trying to recover a U(1) global symmetry by sending a gauge coupling to zero. Given that when a global shift symmetry of a field is broken, the global symmetry of the Hodge dual field is gauged, both conjectures could just be \emph{dual} versions of each other.

This intuition of restoring a global symmetry was obtained in~\cite{Grimm:2018ohb} by studying infinite distance singularities in the complex structure moduli space of type IIB Calabi-Yau compactifications. There, the discrete monodromy transformation generating the infinite tower translates into a discrete shift symmetry of the axionic complex structure modulus corresponding to the angular coordinate encircling the singularity. In this paper, we will show how this intuition can be extrapolated to K\"ahler moduli spaces.  In fact, even if the moduli space is not complex, as M-theory on a Calabi-Yau threefold or the circle compactification of F-theory, it will still be possible to have a notion of a \emph{monodromy transformation} which will generate the tower and will correspond to some $p$-form discrete shift symmetry in the effective theory. In particular, we will see that in M-theory Calabi-Yau threefold compactifications, the discrete symmetry enhances to a continuous one-form global symmetry at infinite distance. This suggests a generalization of the SDC by requiring an infinite number of massless degrees of freedom (not necessarily particles) at every infinite distance point at which a generalized global symmetry would be restored (see \cite{Gaiotto:2014kfa} for a detailed explanation of generalized global symmetries). It would be interesting to further investigate this relation between the SDC and generalized global symmetries in the future.

\subsection{Kaluza-Klein circle compactification}
\label{sec:KK}

As mentioned, the aim of this paper is to study infinite distance limits in the K\"ahler moduli space of a string compactification. The expectation value of the K\"ahler moduli parametrize the volumes of non-trivial cycles of the compactification space. Hence, in certain cases, moving in this moduli space  will also correspond to varying the Planck mass as this is given by the overall volume of the internal space. It is important to remark that the mass of the tower of states in~\eqref{SDC} is given in the Einstein frame, which implies that~$\mp$ is assumed to remain fixed. Otherwise, the mass in~\eqref{SDC} should be replaced by the ratio~$m/\mp$. This implies, in particular, that the tower of states at infinite distance can be very heavy while still satisfying ~$m/\mp\rightarrow 0$ if~$\mp\rightarrow \infty$ at infinite distance. In other words, the tower of states only affects the low energy effective theory if~$\mp$ is finite, but any effect disappears if gravity decouples, as expected from a swampland constraint. The simplest example in which this happens corresponds to varying the radius of a circle compactification. For this reason, we will first describe these observations on a Kaluza-Klein circle compactification as well as the meaning of the species bound in this context, before turning to more complicated K\"ahler moduli spaces in string theory in section~\ref{TypeIIAorbits}.

To begin with,
we consider the effective theory of a complex scalar field in~$D+1$-dimensions,
\begin{align}
S_{D+1} = \mpl{D+1}^{D-1} \int_{\c M_{D+1}} \left \{ \hat{ {R} } + \partial_\mu \bar{ \hat \phi} \,  \partial^\mu \hat \phi\right\} \hat \star\, 1,
\end{align}
and dimensionally reduce it on a circle satisfying~$\d \hat s^2 = \d s^2 + r^2 \d y^2$. Our convention is that hatted objects are~$D+1$-dimensional,~${R}$ is the Ricci scalar and~$\mpl D$ is the~$D$-dimensional Planck mass. A circle has a single modulus~$r$ whose expectation value parametrizes the radius of the circle. The kinetic term for~$r$ only appears after performing the Weyl rescaling~$g^E_{ab} = \big ( \frac{r}{r_0}\big )^{\frac{2}{D-2}} g_{ab}$ to go to the Einstein frame of the~$D$-dimensional theory,
\begin{align}\label{afterWeyl}
S_D = \mpl{D}^{D-2}  \int_{\c M_{D}} \bigg  \{ { {R}^E } + {\textstyle \,\frac{D-1}{D-2}} \frac{1}{r^2} \partial_a r \, \partial^a r+  \sum_{n \in \Z} \left(\partial_a  \bar \phi_n \,  \partial^a \phi_n  + m_n(r)^2 \bar \phi_n \phi_n\right) \bigg  \}  \star^E 1\ ,
\end{align}
where the introduction of the scale~$r_0$ is required to keep the metric dimensionless and can be later fixed to the expectation value of~$r$.
The field metric for~$r$ exhibits  infinite distance singularities at~$r\to 0$ and~$r \to \infty$; the Planck masses in~$D$ and~$D+1$ dimensions are related by
\begin{align} \label{Mp_circle}
\mpl{D}^{D-2} \sim r_0 \, \mpl{D+1}^{D-1}.
\end{align}
The~$D+1$-dimensional scalar field leads to a massless scalar field plus a tower of massive Kaluza-Klein modes of mass~$m_n(r) = \frac{n}{r} \left(\frac{r_0}{r}\right)^{\frac{1}{D-2}}$. This tower of KK modes becomes massless in the decompactification limit~$r\rightarrow \infty$ and their mass decreases exponentially in terms of the proper field distance  $  \Delta  = \a \log r $, where $ \a =  \sqrt{\frac{D-1}{D-2}} $,
\begin{equation}
m_n=n \, r_0^{\frac{1}{D-2}}\exp \big (\!- \a \Delta \big)\ ,\quad\end{equation}
consistent with the Swampland Distance Conjecture.
The species bound~\eqref{SpeciesBound} for the KK tower reads
\begin{align}
\L_{QG} \lesssim \bigg (\frac{M_{\rm pl,D}^{D-2}}{r_0}\bigg 
)^{\frac{1}{D-1}} \left( \frac{r_0}{r}\right)^{\frac 1 {D-2}} \sim \mpl{D+1} \left( \frac{r_0}{r}\right)^{\frac 1 {D-2}},
\end{align}
where we have used that~$\D m = \frac{1}{r} \left(\frac{r_0}{r}\right)^{\frac{1}{D-2}}$.
Therefore, the true quantum gravity cutoff~$\L_{QG}$ is indeed dictated by~$\mpl{D+1}$ and not~$\mpl D$,  which fits with the fact that the UV of the theory is in fact higher dimensional. In other words, for an observer in D-dimensions, the presence of the tower of KK modes lowers the quantum gravity cut-off from~$M_{\rm pl,D}$ to~$\L_{QG}\sim M_{\rm pl,D+1}$ and this matches with the fact that this is also the scale at which quantum gravitational effects become important for an observer in~$D+1$.
The number of species present at this scale is 
\begin{align}\label{N_QG}
N_{QG} \sim r \, \bigg (\frac{\mpl D^{D-2}}{r_0}\bigg)^{\frac{1}{D-1}} \sim  r \, \mpl{D+1} \ .
\end{align}
Notice also that the quantum gravity cut-off~$\L_{QG}$ goes to zero only if one insists on keeping~$M_{\rm pl,D}$ fixed. However, in the usual picture one rather keeps~$M_{\rm pl,D+1}$ fixed so that~$M_{\rm pl,D}$ goes to infinity as~$r\rightarrow \infty$.

We can also compute the quantum corrections from the KK tower to the field metric integrating up to~$N_{QG}$. Notice that this is not a standard regularization method as we want to explicitly keep the dependence on the UV cut-off. Recall that~$\L_{QG}$ depends on~$r$ and this dependence is crucial to generate the infinite field distance. Using~\eqref{goneloop} we obtain
\begin{align}
\delta g^{\oneloop} &\sim \sum_{n=-N_{QG}}^{N_{QG}} m_n(r)^{D-4} \, \partial_\phi m_n(r) ^2 \sim N_{QG}^{D-1} r_0 r^{-D-1}
\sim \mpl{D}^{D-2} \, \frac{1}{r^2} \label{g1L} \,
\end{align}
which has the same parametric dependence as the classical piece in~\eqref{afterWeyl}. Therefore, we find that integrating out the infinite tower of KK modes up to the species bound, one generates a metric that forces the limit~$r\to \infty$ to be at infinite distance. This is expected as it corresponds to a particular case of the general computation performed in the previous section. However, notice that this is a mild version of emergence, as the metric already has a classical divergence.  One could wonder if this classical piece could also emerge from integrating out another infinite tower of states. Even if this is not possible in a Kaluza-Klein compactification, it might be possible in a consistent theory of quantum gravity. We will discuss  this issue again when studying a circle compactification of 6D F-theory in section~\ref{MF-section}. It would also be interesting to study how typical regularization methods applied to UV-dependent quantities change when we assume that the UV cut-off varies. Let us also recall that if we keep~$\L_{QG}$ fixed instead and apply usual regularization methods, we do not get any quantum divergence for the field distance, but in return, the $D$-dimensional Planck mass tends to infinity and gravity decouples. Only if we insist on keeping~$M_{\rm pl,D}$ fixed, we generate the infinite field distance at quantum level.

The possibility of having different towers becoming massless at infinite distance raises new questions: is there any preferred tower that should be identify as the candidate for the SDC? Is it always possible to find a tower responsible for the quantum emergence of the infinite field distance? We think that the best way to identify the tower is to look for the objects that are charged under the discrete symmetry that becomes continuous at infinite distance. And this is what will do in the rest of the paper, by identifying the charge orbit of states generated by a monodromy transformation of infinite order. This monodromy is part of the discrete duality group of the compactification which enhances to a continuous group at the infinite field distance singularities. Sometimes this tower will correspond to KK modes but in general it will consist of more exotic objects, namely wrapping D-branes.

As a final comment, let us recall that the limit~$r\rightarrow 0$ is also at infinite distance. From the point of view of this quantum field theory, there is not any additional tower that become massless in this limit. However, if the theory has a stringy UV-completion, one has indeed the tower of winding modes becoming massless as~$r\to 0$ (this is actually a motivation for a theory of extended objects~\cite{Ooguri2007}, if one assumes the SDC to hold). Even if this limit is usually not accessible in a supergravity effective theory, we can analyze it in the context of string theory, by making use of the T-duality. Since under T-duality winding modes and KK states are exchanged, and that the metric~$\frac 1 {r^2} (\partial r)^2$ is left invariant, one can conclude from the above analysis of integrating out the KK-modes that, at small radius values, the result of integrating out the winding modes will also yield a metric~$\sim \frac 1 {r^2} (\partial r)^2$, thereby also forcing the limit~$r\to 0$ at infinite distance. We will come back to such arguments involving dualities in section~\ref{sec:smallfiber}.

\section{Infinite distances and charge orbits at large volume in Type~IIA} \label{TypeIIAorbits}

In this section we shift to the discussion of the SDC in string theory. More precisely, 
we will consider Type IIA string theory compactified on a Calabi-Yau threefold~$Y_3$. Focusing 
vector multiplet sector of the resulting~$\cN=2$ four-dimensional theory we study infinite 
distances in K\"ahler moduli space. Note that the K\"ahler moduli, henceforth denoted by~$v^I$, 
parameterize the volumes of geometrical submanifolds of~$Y_3$. Limits sending 
one or more~$v^I$ to infinity hence correspond to decompactification limits in generalization 
of the discussion of section~\ref{sec:KK}. We will classify such limits in subsection~\ref{class_infinite_distance_limits}
and show that they always lead to infinite distances in subsection~\ref{infinite_distances_sec}. 
The  candidate charge orbits of states that become massless in the limits are determined in subsection~\ref{Kahler_charge_orbits}. They can be explicitly constructed and studied 
for elliptic fibrations, as we show in subsection~\ref{sec:infinite_elliptic}. Finally, we show in subsection~\ref{sec:smallfiber} 
that in the latter case the orbits can be transferred from large to small elliptic fiber volume. 

\subsection{Classifying infinite distance limits in the large volume regime}\label{class_infinite_distance_limits}

To start with we briefly review some basic aspects of the K\"ahler moduli space of Type IIA Calabi-Yau compactifications.  The moduli space~$\mathcal{M}_{\rm Ks}$ is a K\"ahler 
manifold of complex dimension~$h^{1,1}$, where~$h^{p,q}=\text{dim}(H^{p,q}(Y_3,\mathbb{C}))$ are the Hodge numbers of the Calabi-Yau threefold~$Y_3$. The complexified K\"ahler structure deformations~$t^I$ parametrizing~$\cM_{\rm Ks}$ are given by
\begin{equation}
\label{t_IIA}
B_2+iJ=t^I\omega_I\ , \quad I=1,\dots,h^{1,1}(Y_3)\ ,
\end{equation}
where the~$\omega_I$'s form a basis of the harmonic (1,1)-forms of~$Y_3$,~$B_2=b^I\omega_I$ is the NS 2-form 
and~$J=v^I\omega_I$ is the K\"ahler form, so~$t^I=b^I+iv^I$. 
The K\"ahler potential is given by~$K=-\log 8\cV$ with the overall volume~$\cV$ is defined as 
\begin{align} \label{def-cVIIA}
\c V = \frac{1}{3!} \int_{Y_3 } J\wedge J \wedge J = \frac{1}{3!} \c K_{IJK} v^I v^J v^K \,,
\end{align} 
where the triple intersection numbers are defined as 
\begin{align}\label{intnbIIA}
\c K _{IJK}  = \int_{Y_3} \o_I \wedge \o_J \wedge \o_K\,.
\end{align}
Furthermore it is useful to introduce~$b_I=\frac{1}{24}\int_{Y_3}\omega_I\wedge c_2(Y_3)$, with~$c_2(Y_3 )$ being the second Chern class of the Calabi-Yau threefold.  
The scalars~$t^I$ comprise~$n_v=h^{1,1}(Y_3)$ vector multiplets together with the vectors~$A^I$ coming from expanding the RR three-form~$C_3$ in the same basis
\begin{equation}
C_3=A^I\wedge \omega_I+\dots\ .
	\end{equation}
Note that there is one further vector in the spectrum arising from the dimensional reduction of the 
RR one-form~$C_1$. This additional vector, or rather an appropriate linear combination of all vectors, will 
be part of the gravity multiplet and is often denoted as the graviphoton. 

Let us next introduce the machinery to classify the types of infinite distances 
that appear in the large volume regime~$v^A \gg 1$ of Calabi-Yau compactifications. The basic idea 
is to translate the data specifying the large volume compactification given in~\eqref{intnbIIA},
i.e.~the triple intersection numbers~$\cK_{IJK}$ and the second Chern class~$b_I$, 
into~$h^{1,1}(Y_3)$ so-called log-monodromy matrices~$N_I$ and an anti-symmetric inner product~$\pairing$. 
Together~$N_I$,~$\pairing$ capture all relevant information concerning the metric 
on the scalar field space spanned by the~$t^I$'s. 

To begin with we briefly discuss the construction of a 
monodromy matrix in K\"ahler moduli space by using mirror symmetry.   
More precisely, recall that under mirror symmetry the large volume point is mapped to the large complex structure point by identifying the complexified K\"ahler structure deformations~$t^I$ with the complex structure deformations~$z^I$ of IIA and IIB compactifications. The K\"ahler potential for complex structure moduli space of the mirror Calabi-Yau threefold~$\tilde Y_3$ is given by
\begin{equation}
\label{Kcs}
K(z,\bar z)=-\log(i\bar \Pi^\cI\pairing_{\cI\cJ}\Pi^\cJ)
\end{equation}
where~$\Pi^\cI$ are the periods of the holomorphic (3,0)-form~$\Omega$ into a real integral basis~$\gamma_\cI$,~$\cI=1,\dots,h^{2,1}(\tilde Y_3)+2$ of three-cycles as follows, 
\begin{equation}
\label{Pics}
\Omega=\Pi^\cI\gamma_\cI\ , \quad \pairing_{\cI\cJ}=-\int_{\tilde Y_3}\gamma_\cI\wedge \gamma_\cJ\ .
\end{equation}
The mirror map implies that, at the large volume point, one can introduce the following~$2h^{1,1}(Y_3)+2$-dimensional period vector~$\mathbf{\Pi}$ 
depending on these complex variables 
\begin{equation}\label{tI-periods}
\mathbf \Pi \, (t^I) =
\begin{pmatrix}
1\\
t^I\\
\frac{1}{2}\cK_{IJK} t^J t^K + \frac{1}{2}\cK_{IJJ} t^J - b_I  \\
\frac{1}{6}\cK_{IJK} t^I t^J t^K - (\frac{1}{6} \cK_{III} + b_I) t^I + \frac{i \zeta(3) \chi}{8\pi^3}  
\end{pmatrix},
\end{equation}
where~$\chi  = \int_{ Y_3} c_3( Y_3)$ is the Euler number of~$ Y_3$. It is crucial in this 
identification that we consider a basis~$\omega_I$ spanning (part of) the K\"ahler cone. In other 
words, we need to ensure that when taking~$v^I > 0$, the K\"ahler form~$J = v^I \omega_I$ measures a 
positive volume~$\int_C J >0$ for all irreducible proper curves~$C$ in~$Y_3$. 
While much of the following discussion is general, we will assume that the K\"ahler cone of the 
considered manifold is simplicial, i.e.~spanned by exactly~$h^{1,1}(Y_3)$ generators. This implies, 
in particular, that all~$\cK_{IJK} \geq 0$, which will significantly simply the discussion below. 

Then one defines the monodromy transformation to be the matrix arising in the 
transformation~$ \mathbf{\Pi} (t^1,\ldots,t^A-1,\ldots ) = T_A \, \mathbf{\Pi}(t^1,\ldots,t^{A},\ldots)$. From the point of view of the four dimensional effective theory, this transformation corresponds to a discrete shift of the axionic field $\text{Re}(t^I)$.
Instead of displaying the matrix~$T_A$ (see~\cite{Grimm:2018cpv} for an explicit expression and references), we rather 
show the nilpotent matrix~$N_A$ obtained from~$T_A$ by setting 
\begin{equation}\label{N=logT}
N_A = \log (T_A)\ . 
\end{equation}
These~$N_A$ are known as the log-monodromy matrices and can be used to classify 
singularity types arising in Calabi-Yau moduli spaces. For the large complex structure periods 
~\eqref{tI-periods} they are readily determined to be
\begin{equation}\label{lcsNa} 
N_A  = \left(
\begin{array}{cccc}
0                   & 0                  & 0            & 0\\
-\delta_{AI}        & 0                  & 0            & 0\\
-\frac{1}{2} \cK_{AAI} & -\cK_{AIJ}           & 0            & 0\\
\frac{1}{6}\cK_{AAA}  & \frac{1}{2}\cK_{AJJ} & -\delta_{AJ} & 0
\end{array}\right)\ . 
\end{equation}
The corresponding pairing~$\pairing$ that can be used to contract the periods, takes the form
\begin{equation} \label{eta-lcs}
\pairing = \left(
\begin{array}{cccc}
0                         & -\frac{1}{6} \cK_{JJJ} - 2b_J     & 0                & -1\\
\frac{1}{6}\cK_{III} + 2b_I & \frac{1}{2}(\cK_{IIJ} - \cK_{IJJ}) & \delta_{IJ}      &  0\\
0                         & -\delta_{IJ}                   & 0                &  0\\
1                         & 0                              & 0                &  0
\end{array}\right)\ .
\end{equation}
where~$b_I$ was introduced below~\eqref{intnbIIA}. 
It is important to stress that displayed~$(2h^{1,1}(Y_3) +2)\times (2h^{1,1}(Y_3) +2)$  matrices~$N_A$,~$\pairing$ 
are determined in a special basis of even 
forms on the Calabi-Yau manifold~$Y_3$, which also requires the K\"ahler cone condition 
introduced above. We will not go into details how this basis is derived, 
but rather stress that the following considerations are invariant under basis transformations. Let us also notice that the above nilpotent matrix has also been derived in a different avenue by analysing the structure of the flux induced scalar potential when written in terms of Minkowski 3-form fields \cite{Bielleman:2015ina,Carta:2016ynn,Herraez:2018vae}, as it also deeply relies on the presence of the discrete axionic shift symmetries.

The crucial point is that we can now associate a log-monodromy matrix to each limit of the~$t^I$ taken in 
the K\"ahler cone. The simplest situation is to consider only a specific~$t^I$ taken to~$i \infty$ for some 
chosen index~$I$. Let us relabel the coordinates such that this is the direction~$t^1$. 
Then one has to associate the matrix~$N_1$ to this limit. 
However, if one takes the limit 
in two directions, which we choose after relabeling to be~$t^{1} \rightarrow i \infty$ and~$t^{2} \rightarrow i \infty$, then one associates the matrix 
$N_{1} + N_{2}$, or any other positive linear combination of~$N_{1}$,~$N_{2}$, to this limit. 
In general, if one takes the limit of~$n$ coordinates labeled by~$t^1,\ldots,t^n$, one thus 
associates 
\begin{equation}\label{limit-map}
t^{1},\ldots,t^{n} \rightarrow i \infty \quad \longrightarrow \quad N_{(n)} = N_{1} + \ldots+ N_{n}\ ,
\end{equation}
where~$N_{(n)}$ is the relevant log-monodromy matrix in this limit. 
For future reference, we give here its explicit form in terms of the intersection numbers
\begin{equation}\label{Nn}
N_{(n)}  = \left(
\begin{array}{cccc}
0               &             0             &         0          & 0 \\
-\sum_i^n\delta_{iI}      &             0             &         0          & 0 \\
-\frac{1}{2} \sum_i^n\cK_{iiI} &       -\sum_i^n\K{iIJ}       &         0          & 0 \\
\frac{1}{6}\sum_i^n\cK_{iii}  & \frac{1}{2} \, \sum_i^n\K{iJJ} & -\sum_i^n\delta_{iJ} & 0
\end{array}\right) .
\end{equation}
Note that in order to extract the 
crucial properties of the limit one can also replace the above~$N_{(n)}$ with any 
other linear combination of all~$N_{1},\ldots,N_{n}$ with positive coefficient. 
The crucial point about this map is the fact that one now has a handle on classifying 
infinite distances by analyzing the associated log-monodromy matrix~\cite{Grimm:2018cpv,Grimm:2018ohb} 
In fact, since the allowed log-monodromy matrices can be classified~\cite{Kerr2017} one also finds a classification 
of limits in the K\"ahler cone and of infinite distances. 

Let us briefly introduce the general classification of log-monodromy matrices for Calabi-Yau threefolds. 
In general these do not have to arise from the large volume regime, even though we will immediately return to this 
specific situation after this brief general interlude. More precisely, they can arise at any limit 
in an~$m$-dimensional complex structure moduli space, of which the large volume regime is 
just a single patch identified via mirror symmetry. 
Let us denote a log-monodromy matrix by~$N$ 
and the inner product by~$\pairing$. The allowed pairs~$(N,\pairing)$ can be classified into~$4 m$ types 
denoted by 
\begin{equation}\label{sing_types}
\begin{alignedat}{2} 
&\text{I}_a\ ,  &&\quad a=0,\ldots,m\ ,  \\
& \text{II}_b\ ,&&\quad b=0,\ldots,m-1 \ ,\\  
&\text{III}_c\ ,&&\quad c= 0,\ldots,m-2 \ , \\ 
&\text{IV}_d\ , && \quad d = 1,\ldots,m \ .  
\end{alignedat}
\end{equation}
In fact, these types classify singularities that can arise at the boundaries of the moduli space. 
Near such a boundary one can introduce local coordinates~$t^I$, and the limits are taken 
as above in~\eqref{limit-map}. The singularity types are distinguished~\cite{Grimm:2018cpv} by the relations displayed in Table~\ref{table1}, where we included the extra condition allowing us to distinguish the 
cases I$_a$ and II$_b$ by using only~$\pairing$ and~$N$.
\begin{table}[!ht]
	\centering
	\begin{tabular}{@{}l@{\hspace{20pt}}lcc@{\hspace{20pt}}r@{}}\toprule
		\multirow{2}{*}{Type} \hspace*{.5cm} & \multicolumn{3}{c}{\hspace*{-2.5em} rank of} & \multirow{2}{*}{eigenvalues of$\pairing N$}  \\[1pt] 
		&$N$ &$N^2$&$N^3$ & \\ \midrule
		I$_a$ &$a$ & 0& 0 &$a$ negative\\
		II$_b$ &$2+b$ & 0& 0 & 2 positive, $b$ negative\\
		III$_c$ &$4+c$ & 2& 0 & not needed\\
		IV$_d$ &$2+d$ & 2& 1 & not needed\\
		\bottomrule
	\end{tabular}
	\caption{Classification of the arising limits and singularities occurring in the complex 
		moduli space of Calabi-Yau threefolds.}
	\label{table1}
\end{table}

Let us stress that the~$N$ appearing in Table~\ref{table1} does not have to be the log-monodromy 
matrix arising from sending a single coordinate into a limit. Rather, it can be extracted when sending 
any number of coordinates~$t^{I}$  to~$i \infty$ as in~\eqref{limit-map}. Hence, we can also study what 
happens if we send step-wise one after the other coordinate to~$i \infty$. 
At the~$j$th step we can determine the singularity type by 
associating the appropriate~$N_{(j)}$ using~\eqref{limit-map}, i.e.~we consider 
\begin{equation}\label{N(i)}
t^{1},\ldots,t^{j} \rightarrow i \infty \quad \longrightarrow \quad N_{(j)} = N_1 + \ldots + N_j\ ,\qquad j=1,\ldots,n\ , 
\end{equation}
and then determined the type using Table~\ref{table1}. As a place-holder for the possible types~\eqref{sing_types} we will 
write~${\sf Type\ A}_{ (j)}$ for the singularity type occurring at the~$j$th step. 
We then find an enhancement chain of the form
\begin{equation}\label{t>inf_chain}
\xrightarrow{\ t^1 \rightarrow i\infty\ }\  {{\sf Type\ A}_{(1)}}\ \xrightarrow{\ t^2 \rightarrow i\infty\ }\  {\sf Type\ A}_{(2)} \ \xrightarrow{\ t^3 \rightarrow i\infty\ }\  
\ldots \ \xrightarrow{\ t^{n} \rightarrow i\infty\ }\  {\sf Type\ A}_{(n)} \ .
\end{equation}
In fact, one can show that the type only can 
increase or stay the same, i.e.~a general chain of singularity enhancements takes the form 
\begin{align}
\begin{split}
 \text{I}_{a_1}\, \rightarrow \ldots  \rightarrow\, \text{I}_{a_k} \, \rightarrow \, \text{II}_{b_1}\, \rightarrow \ldots  \rightarrow\, \text{II}_{b_l}\,  \rightarrow \hspace*{4cm}
\\
 \, \rightarrow \, \text{III}_{c_1}\, \rightarrow \ldots  \rightarrow\, \text{III}_{c_p} \, \rightarrow \, \text{IV}_{d_1}\, \rightarrow \ldots  \rightarrow\, \text{IV}_{d_q}\ . 
\end{split}
\end{align}

The precise rules of which enhancements can occur in principle have been worked out in~\cite{Kerr2017} and 
a concise summary can be found in~\cite{Grimm:2018cpv}, Table 3.3.

This general classification can immediately be applied to the large volume log-monodromies
determined in~\eqref{lcsNa} in the limit~$t^A \rightarrow i \infty$ for a single coordinate~$t^A$.
In this case it is not hard to show by using~\eqref{lcsNa},~\eqref{eta-lcs} together with the fact 
that~$\cK_{IJK}\geq 0$ for a simplicial K\"ahler cone, 
that the case I$_a$ actually does not arise in the large volume regime~$\I\, t^I \gg 1$. 
This matches the fact that the Type I$_a$ corresponds to having a finite distance in moduli 
space. It arises, for example, at the conifold point in complex structure moduli space, but not at 
the large volume regime where all limits are expected to be at infinite distance. 
For the remaining three cases, the singularity type of the individual limits 
$t^A \rightarrow i \infty$ is evaluated by considering~$N_A$ given in~\eqref{lcsNa}, an first computing its square and cube 
\begin{equation}       
N_{A}^2  
= \left(
\begin{array}{cccc}
0    &    0    & 0 & 0 \\
0    &    0    & 0 & 0 \\
\K{AAI} &    0    & 0 & 0 \\
0    & \K{AAJ} & 0 & 0
\end{array}\right), \qquad
N_A^3  
= \left(
\begin{array}{cccc}
0     & 0 & 0 & 0 \\
0     & 0 & 0 & 0 \\
0     & 0 & 0 & 0 \\
- \K{AAA} & 0 & 0 & 0
\end{array}\right) .
\end{equation}
Then, one can evaluate the ranks of~$N_A$,~$N_A^2$ and~$N_A^3$ and use table~\ref{table1} in order to determine the singularity types. The results are summarized in Table~\ref{Type_Table}. 
\begin{table}[!ht]
	\centering
	\begin{tabular}{@{}lccr@{}} \toprule
		Type &~$\rk (\cK_{AAA})$ &~$\rk (\cK_{AAI})$ &~$\rk (\cK_{AIJ})$ \\ \midrule 
		II$_b$ & 0 & 0 &~$b$ \\  
		III$_c$ & 0 & 1 &~$c+2$ \\
		IV$_d$ & 1 & 1 &~$d$ \\ \bottomrule
	\end{tabular}
	\caption{We list the singularity types arising in the large volume regime, 
		when sending a single coordinate~$t^A \rightarrow i \infty$.
		Note that that the ranks~$\text{rk}(\cK_{AAA})$ and~$\text{rk}(\cK_{AAI})$ are either~$0$ or~$1$ depending on whether  
		$\cK_{AAA}$ and~$\cK_{AAI}$ are vanishing or not.}
	\label{Type_Table}
\end{table}

These results can be straightforwardly generalized to the case of sending multiple~$t^I$ to~$i\infty$ in K\"ahler moduli space. As before we relabel the coordinates such that the limit of interest sends the first~$n$ coordinates to~$i \infty$.
The relevant log-monodromy matrix associated to this limit is~$N_{(n)}$, introduced in~\eqref{Nn}. 
Introducing the notation
\begin{equation}\label{notation}
\K{IJ}^\np \equiv \sum_{i=1}^n \K{iIJ}, \qquad \K{I}^\np \equiv \sum_{i,j=1}^n \K{ijI}  \qquad \text{and} \qquad \K{}^\np \equiv \sum_{i,j,k=1}^n \K{ijk}\,,
\end{equation}
we find
\begin{equation}\label{Nn2}
N_{(n)}^2 =  \left(
\begin{array}{cccc}
0     &     0     & 0 & 0 \\
0     &     0     & 0 & 0 \\
\K{I}^\np &     0     & 0 & 0 \\
0     & \K{J}^\np & 0 & 0
\end{array}\right), \quad 
N_{(n)}^3  
= \left(
\begin{array}{cccc}
0      & 0 & 0 & 0 \\
0      & 0 & 0 & 0 \\
0      & 0 & 0 & 0 \\
- \K{}^\np & 0 & 0 & 0
\end{array}\right).
\end{equation}
Evaluating the ranks of~$N_{(n)}$,~$N^2_{(n)}$ and~$N^3_{(n)}$ and using again Table~\ref{table1}, one finds the singularity type. This yields a generalization of Table~\ref{Type_Table}, which is presented in Table~\ref{Type_Table2}.
\begin{table}[!ht]
	\centering
	\begin{tabular}{@{}lccr@{}} \toprule
		Type &~$\rk \cK_{}^\np$ &~$\rk \cK_{I}^\np$ &~$\rk \cK_{IJ}^\np$ \\ \midrule 
		II$_b$ & 0 & 0 &~$b$ \\  
		III$_c$ & 0 & 1 &~$c+2$ \\
		IV$_d$ & 1 & 1 &~$d$ \\ \bottomrule
	\end{tabular}
	\caption{We list the singularity types arising in the large volume regime, 
		when sending multiple coordinates~$t^{1},\ldots,t^{n} \rightarrow i \infty$.
		Note that that the ranks~$\text{rk}(\text{number})$ and~$\text{rk}(\text{vector})$ are either~$0$ or~$1$ depending on
		whether the number and vector are vanishing or not.}
	\label{Type_Table2}
\end{table}

\subsection{Infinite distances in K\"ahler moduli space} \label{infinite_distances_sec}

Having classified the limits in K\"ahler moduli space we next 
study the distances along paths as measured by the K\"ahler metric~$K_{I\bar J} = \partial_{t^I} \partial_{\bar t^J} K$. 
Recall that the length of a path connecting two points~$Q,P$ in moduli space is 
determined by  
\begin{equation}\label{d(P,Q)}
d_\gamma(Q,P) = \int_\gamma \sqrt{2\, K_{I\bar J} \ \dot t^I \dot{\bar{t}}^J}\, ds\ ,
\end{equation}
where the path~$\gamma$ is parameterized in local coordinates by~$t^I(s)$ 
and we abbreviated~$\dot t^I = \frac{\partial t^I}{ \partial s}$.
In the following we will show that each path approaching a point~$P$ that is located at  
$t^1,\ldots,t^n \rightarrow i \infty$, for some~$n$, is infinitely long.

To begin with, we determine the K\"ahler potential using~\eqref{Kcs} and inserting 
the mirror periods~$\Pi$ given in~\eqref{tI-periods} and the intersection form~$\pairing$ given 
in~\eqref{eta-lcs}. This yields the well-known expression
\begin{equation}\label{Klarge_volume}
K = - \text{log} \Big(\frac{1}{6}\cK_{IJK} v^I v^J v^K +\frac{\zeta(3) \chi}{32\pi^3} \Big) \equiv - \text{log}\, \cV_{\rm q} \ . 
\end{equation}
Clearly, if we consider simplicial K\"ahler cones, we can use~$\cK_{IJK} \geq 0$ to infer that~$\cV_{\rm q}$ 
diverges and hence
$K$ approaches negative infinity for any limit~$v^1,\ldots,v^n \rightarrow  \infty$. If we want to 
work more generally and also want to infer the growth of~$\cV_{\rm q}$, we can apply 
a result determined in~\cite{Grimm:2018cpv} based on the growth theorem of~\cite{CKS}. 
More precisely, one shows that the leading growth of~$\cV_{\rm q}$ is 
\begin{equation}\label{eKgrowth}
\cV_{\rm q}\sim  c\, \left( v^1 \right)^{d_1} \left( v^2 \right)^{d_2-d_1} \cdots(v^{n})^{d_{n} - d_{n-1}}  \ . 
\ , 
\end{equation}
if one considers the limit~$v^1,\ldots,v^n \rightarrow \infty$ in the growth sector
\begin{equation}\label{sector}
\left\{ \frac{v^{1} }{v^2 } > \lambda \, ,\ldots , \ 
\frac{v^{{n}-1}}{v^{{n}} } > \lambda  \,  ,\   v^{n} > \lambda\,  \right\} 
\ ,
\end{equation}
for some positive~$\lambda$. Here~$c$ is a positive constant and the symbol~$\sim$ indicates 
that we only focus on the leading term. The integers~$d_i$ are simply the types occurring 
in the corresponding enhancement chain~\eqref{t>inf_chain}, i.e.~we identify 
\begin{equation}\label{associated_di}
\begin{tabular}{@{}llcr@{}}
\toprule
$\mathsf{Type\ A}_{(i)}$ & II$_b$ & III$_c$ & IV$_d$ \\
\midrule 
$d_i$ & 1 & 2 & 3\\
\bottomrule
\end{tabular}
\end{equation}
With this identification it is now clear that maximally three~$v^i$ can appear 
in~\eqref{eKgrowth} as expected from~\eqref{Klarge_volume}.
It is crucial to point out that the growth of~$\cV_{\rm q}$ depends on 
the sector~\eqref{sector} into which that path~$\gamma$ towards~$P$ falls. 
This is not a very restrictive constraint on the considered paths, since one 
can reorder the~$v^1,\ldots,v^n$ to satisfy the inequalities in~\eqref{sector}. 
Accordingly one then has to also consider an appropriately reordered 
enhancement chain~\eqref{t>inf_chain} and adjust the~$d_i$. 

Having determined the growth of the K\"ahler potential, let us now determine the 
growth of the length of the path. In the following we will establish a lower bound on 
this growth. To do that, let us first assume 
\begin{equation}\label{ns-bound}
2 K_{I} K^{I\bar J} K_{\bar J} \leq f^{-2}  \ , 
\end{equation}
where~$f$ is some constant that might depend on the choice of the path~$\gamma$.
We will show below that this condition is indeed satisfied for the K\"ahler potential~\eqref{Klarge_volume}. 
In order to find a lower bound on the length of a cure we use~\eqref{ns-bound} and the Cauchy-Schwarz inequality  
to derive \footnote{The Cauchy-Schwarz inequality reads~$||v || \cdot ||u|| \geq | \langle v , u\rangle|$, where the norm is 
	related to the inner product by~$|| v|| = \sqrt{\langle v , v\rangle}$. In the case at hand one uses~$v \cong ( \dot t^I , \dot{\bar{t}}^J)$,~$u \cong ( K^{I\bar L} K_{\bar L} , K^{\bar J L} K_L)$, with an inner product determined by the K\"ahler metric. } 
\begin{equation}
(2 K_{I \bar J} \dot t^I \dot{\bar{t}}^J)^{1/2} \geq f \,(2 K_{I\bar J} \dot t^I \dot{\bar t}^J)^{1/2} (2 K_{I} K^{I \bar J} K_{\bar J})^{1/2} 
\geq  f\, |K_{I} \dot t^I + K_{\bar J} \dot{\bar{t}}^J|= f\, |\dot K |
\end{equation}
Using this estimate in~\eqref{d(P,Q)} we find 
\begin{equation}
d_\gamma(Q,P) \geq f \int_\gamma |\dot K|  ds \geq f \Big|\int_\gamma dK  \Big|\ .
\end{equation}
We can integrate the last integral to evaluate 
\begin{equation}
d_\gamma(Q,P) \geq  f  |K(P) - K(Q)|\ ,
\end{equation}
where~$K(P),\ K(Q)$ is the K\"ahler potential evaluated at the two endpoints~$P$,~$Q$. 
This implies that for a point~$P$ with~$t^1,\ldots,t^n \rightarrow i \infty$ we see that the growth of~$ d_\gamma(Q,P)$
is dominated by the divergent contribution near~$P$. Hence, we find that the growth of the length is dominated by
\begin{equation}
d_\gamma(Q,P) \gtrsim f \ \text{log}\,\cV_{\rm q} 
\sim f\,\sum_{i=1}^n ( d_{i} - d_{i-1})\ \text{log}\, v^i \ , 
\end{equation}
with~$d_0 = 0$, and~$d_i$,~$i=1,\ldots,n$ defined in~\eqref{associated_di}. Here we have used the expression~\eqref{eKgrowth}
for the growth of~$\cV_{\rm q}$ in a growth sector~\eqref{sector}.
Clearly, this implies that the length is infinite 
as soon as we take~$v^1,\ldots,v^n \rightarrow \infty$. Note, however, that this does not necessarily imply that every 
path has a length growing logarithmically in~$v^i$, since we only presented a lower bound. 

It remains to show that~\eqref{ns-bound} is actually satisfied for the K\"ahler potential~\eqref{Klarge_volume}. 
By a straightforward computation one determines
\begin{equation}
2 K_{I} K^{I\bar J} K_{\bar J} =  6+6 \sum_{n=1}^{\infty} \left(\frac{\zeta(3) \chi}{16\pi^3 \cK_{IJK} v^I v^J v^K}\right)^n \,.
\end{equation}
Since the non-constant terms are increasingly suppressed in approaching the point~$P$, this implies that one can easily find a constant~$f$ such that~\eqref{ns-bound} is satisfied. 

We thus conclude that all limits in the large volume regime are at infinite 
distance. While this result is not unexpected, it is satisfying to see that it can be 
explicitly derived. It implies that one cannot find finite lengths paths towards 
$t^1,\ldots,t^n \rightarrow i \infty$ by using seemingly appearing  
cancellations in the volume~$\cV_{\rm q}$ due to a choice of basis or a consideration 
of non-simplicial K\"ahler cones. 
It also gives further evidence that limits are at infinite distance if 
and only if the arising singularity types are II, III, or IV. These are precisely the types that 
arise in the large volume regime, as discussed in subsection~\ref{class_infinite_distance_limits}. Note that 
only the direction that infinite distance implies type II, III, IV singularities has been proved generally 
in a multi-dimensional moduli space~\cite{Wang97onthe}.

\subsection{Infinite charge orbits of states } \label{Kahler_charge_orbits}

Having determined the possible infinite distance singularities 
arsing in the large volume regime, 
we next want to identify an infinite set of states that become massless when 
approaching such limits.
It was suggested in~\cite{Grimm:2018ohb} that these states are 
generated by acting with the monodromy matrix on a single seed charge~$\mathbf{q}_0$
to generate an infinite tower. In higher-dimensional field spaces this can be captured 
by what was called a charge orbit denoted by~$  \mathbf{Q}(\mathbf{q}_0 | m_1,\ldots,m_k)$ in~\cite{Grimm:2018cpv}. 
There are two basic requirements on the charge orbit~$\mathbf{Q}(\mathbf{q}_0 | m_1,\ldots,m_k)$ for it to generate
the states necessary in the SDC. Firstly, the 
states have to become massless when approaching an infinite distance point. Secondly, 
there has to be infinitely many states with this feature. It was suggested in~\cite{Grimm:2018ohb}
that such states are actually BPS states with mass determined by the central charge 
$M(\mathbf{Q}) = |Z(\mathbf{Q})|$. The tricky part of this study is to evaluate 
the behavior of~$M(\mathbf{Q})$ along every path approaching the infinite distance 
point. This can be done by splitting the moduli space near the infinite distance points 
into growth sectors as we discuss in the following.

To begin with we have to determine the growth sector in which a 
given path~$\gamma$ towards a point~$P$ with~$t^1,\ldots,t^n \rightarrow i \infty$ lies. 
A general path can be parameterized by local coordinates~$t^I(s)$, where~$s$ labels  
the position on~$\gamma$. 
To check the growth sector into which~$t^I(s)$ falls, we first introduce it 
for a specific ordering~$t^1,\ldots,t^n$. In this simplest situation it takes the form 
\begin{equation}\label{growth-sector}
\cR_{1\ldots n} \equiv \bigg \{t^i = b^i + i v^i: \quad \frac{v^{1} }{v^2 } > \lambda \, ,\ldots , \ 
\frac{v^{n-1}}{v^{n} } > \lambda  \,  ,\  v^{n} > \lambda\, ,
\quad |b^i| < \delta\, \bigg \} ,
\end{equation}
for some positive~$\lambda,\delta$. It might be the case that this condition cannot be satisfied for~$t^I(s)$
even if we start with very large~$v^i$. Then we have to reorder the~$t^i$ by also exchanging the~$v^i$ in~\eqref{growth-sector}.
Once we have determined an appropriate ordering, we get an order~$(t^{i_1},\ldots,t^{i_n})$ for performing the 
limit~$t^1,\ldots,t^n\rightarrow i \infty$. For this ordering one then has to determine the singularity chain
\begin{equation} \label{enh_chain_2}
\xrightarrow{\ t^{i_1} \rightarrow i\infty\ }\  {{\sf Type\ A}_{(1)}}\ \xrightarrow{\ t^{i_2} \rightarrow i\infty\ }\  {\sf Type\ A}_{(2)} \ \xrightarrow{\ t^{i_3} \rightarrow i\infty\ }\  
\ldots \ \xrightarrow{\ t^{i_n} \rightarrow i\infty\ }\  {\sf Type\ A}_{(n)} \ .
\end{equation} 
Clearly, we can always relabel the coordinates~$t^i$ to make the singularity chain look like~\eqref{t>inf_chain}
and the growth sector takes the form~\eqref{growth-sector}. In the 
following we will assume that such a reordering and relabeling has been performed if necessary. 

Having identified a growth sector and an associated enhancement chain we next want to determine the 
charge orbits relevant in the large volume regime. Later on we will apply this construction to elliptic 
fibrations. Let us first note that there are~$h^{1,1}(Y_3)$ log-monodromy matrices~$N_I$ arising 
in the large volume regime. Each is associated to a coordinate~$t^I$ as discussed above. 
Hence, we expect the general charge orbit to be of the form~\cite{Grimm:2018cpv}
\begin{equation}\label{Q-N}
\mathbf{Q}\big (\mathbf{q}_0 | m_1,\ldots,m_{h^{1,1}(Y_3)} \big ) = \text{exp} \Big ( \textstyle \sum_{I=1}^{h^{1,1}(Y_3)} m_I N_I \Big ) \, \mathbf{q}_0\ ,   
\end{equation}
where we take the~$m_I$'s to be non-negative integers.
Note that this expression simply states that we apply~$m_I$ times the 
monodromy transformations~$T_I$ discussed before~\eqref{N=logT} to a suitable seed charge~$\mathbf{q}_0$. 
The challenge is now two-fold:  (1) one needs to construct a suitable~$\mathbf{q}_0$, which ensures that~$\mathbf{q}_0$ and~$\mathbf{Q}$ are massless at~$P$; (2) one needs to identify situations when~$\mathbf{Q}$ describes an infinite set of states. 
Both of these issues have been clarified in~\cite{Grimm:2018cpv}. However, it should be stressed that the 
explicit constructions of~\cite{Grimm:2018cpv} uses a significant amount of mathematical technology related to 
the construction of a special 
set of matrices~$N_I^-$ that are parts of commuting~$\mathfrak{sl}(2)$ algebras. While in this picture the existence and properties of~$\mathbf{q}_0$ and~$\mathbf{Q}$ can be more easily abstractly analyzed, it is technically involved to construct these special~$N_I^-$.
We will therefore follow a different route here. We will use the conditions found in~\cite{Grimm:2018cpv} translated to the~$N_I$ basis 
and construct the~$\mathbf{q}_0$ satisfying them. Let us stress that the construction of~$\mathbf{q}_0$ is not generally 
expected to be unique and there can be various different charge orbits labeling the relevant states for the SDC.

In reference~\cite{Grimm:2018cpv} it was shown that there are three singularity patterns for which 
generally an infinite charge orbit exists that becomes massless at the considered point 
$P$. The first possibility is that~$P$ lies on a Type IV locus. In other words,~$\mathsf{Type\ A}_{(n)} = \text{IV}$ in the enhancement chain~\eqref{t>inf_chain}. 
The second possibility is that~$P$ lies at a Type II locus, i.e.~that~$\mathsf{Type\ A}_{(n)} = \text{II}$ 
in~\eqref{t>inf_chain} and along this locus occurs an enhancement II~$\rightarrow$ III or II~$\rightarrow$ IV 
in the considered region of field space. Finally, the third possibility is that~$P$ lies at a Type III locus, i.e.~$\mathsf{Type\ A}_{(n)} = \text{III}$ and this singularity enhances as  III~$\rightarrow$ IV in the considered region of field space.
In the large volume regime one of these three possibilities
is satisfied for every infinite distance point~$P$~\cite{Grimm:2018cpv}.  
This can be deduced from the fact that the highest singularity type in the large volume regime is IV$_{h^{1,1}(Y_3)}$. 
Hence, either one is directly at a type IV singularity or one inherits the orbit from the large volume 
point with  IV$_{h^{1,1}(Y_3)}$.
This implies that at each infinite distance point in the large volume regime 
there exists an infinite charge orbit.  
Notice that, if these intersections of the singular divisors allowing the enhancement of the type of singularity had not be present, there would not be possible to identify an infinite charge orbit at type II and III singularities. This exemplifies how the Swampland Distance Conjecture is realized in a highly non-trivial and intricate way in Calabi-Yau compactifications. The conjecture does not constrain only the local structure of the Calabi-Yau but also the global network of enhanced singularities allowed in the moduli space. 

\paragraph{Masslessness conditions}
To construct a charge orbit relevant in the large volume regime let us recall that, since~\cite{Grimm:2018cpv}
\begin{equation}
m(\mathbf Q) = \vabs{Z(\mathbf Q)} \leq \norm{\mathbf Q} \sim \norm{\qo} \,,
\end{equation}
a sufficient condition to ensure the masslessness of~$\mathbf Q$ is that $\qo$ has vanishing norm.
In order to achieve this 
for the enhancement chain~\eqref{t>inf_chain} within the growth sector~\eqref{growth-sector}, we first require that for every $i=1,\ldots,n$ there exists some vectors $\u_i$, $\v_i$ and $\x_i$ satisfying $\N i^2 \v_i =0$ and $\N i^3 \x_i =0$ such that the seed vector takes the form ~\cite{Grimm:2018cpv}
\begin{subequations} \label{masslessness}
	\begin{alignat}{2} \label{masslessness1}
	 \qo & = \v_i &&\qqtext{if} \type{i} = \rII \ ,\\
 	 \qo & = \v_i + \N{i} \u_i && \qqtext{if} \type{i} = \rIII \ ,\\
 	 \qo & = \v_i + \N i \x_i  && \qqtext{if} \type{i} = \rIV \ . 
	\end{alignat}
\end{subequations}
These conditions might not be strong enough to generally ensure masslessness, since in some cases $\norm{\qo}$ might tend to a finite value at the infinite distance point. To make sure that this does not happen, we need to additionally require that for the last singularity in the chain~\eqref{t>inf_chain} that ther exists some vectors $\u_n$ and $\w_n$ satisfying $\N n^2 \w_n = 0 $ such that 
\begin{subequations}\label{masslessnessII}
	\begin{alignat}{2}
\qo & = \N n \u_n         && \qqtext{if} \type{n} = \rII \ ,\\
\qo & = \N n \w_n        && \qqtext{if} \type{n} = \rIII \ ,\\
\qo & = \N n \w_n + \N n ^2 \u_n  && \qqtext{if} \type{n} = \rIV \ .
\end{alignat}
\end{subequations}
Roughly speaking the 
conditions~\eqref{masslessnessII} 
ensure the necessary suppression of $\norm{\qo}$
by at least one coordinate~$v^n$ that grows to infinity 
at the infinite distance point.\footnote{This can be shown by using the results of section 4.3 of reference~\cite{Grimm:2018cpv}. 
	A sufficient condition for~$\mathbf{q}_0$ to be masslessness at the infinite distance point 
	was given in eq.~(4.29). Replacing~$N^-_{(i)} \rightarrow N_{(i)}$
	the condition (4.29) of~\cite{Grimm:2018cpv} is satisfied when imposing~\eqref{masslessness} and~\eqref{masslessnessII}.} 

\paragraph{Infiniteness conditions}
In order to assure that an orbit generated, we need to demand that the action of the exponential in~\eqref{Q-N} on~$\qo$ is non-trivial, i.e.~we need that 
\begin{equation}\label{cond_orbit}
N_{(J^*)} \qo \neq 0 \qqtext{for some} J^*=1,\ldots,h^{1,1}(Y_3) \,.
\end{equation}
Notice that it is enough if this is satisfied for at least one~$N_{J^*}$,~$J^*=1,\ldots,h^{1,1}(Y_3)$. 
In light of the masslessness conditions~\eqref{masslessness}, a simple way how to realize this might be to demand that it is satisfied for a Type IV singularity.
As mentioned above, the large volume point is a Type IV singularity, so we are ensured that an infinite orbit can be generated, even if~${\sf Type\ A}_{(n)} \ne \rIV$. 
However, let us remark that \textsl{stricto sensu} one does not need to have a Type IV singularity in order to generate an orbit. Indeed what can happen is that in a sufficiently small neighborhood~$\c E$ 
close to the point of interest~$P$, there is another Type II or III singularity, associated to a coordinate~$t^J$ which is not taken to~$i\infty$, i.e.~$J>n$. We do not need to impose~\eqref{masslessness1} for~$N_{(J)}$, such that we can have~$N_{(J)}\qo \neq 0$, generating an orbit. Notice that the sum in~\eqref{Q-N} should really be only over the~$N_I$'s present in~$\c E$. 

\paragraph{Constructing the orbit}
We now construct explicitly such a seed vector~$\mathbf{q}_0$. 
We first split this~$2h^{1,1}(Y_3)+2$-dimensional vector into four parts 
\begin{equation}\label{ansatz_q0}
\mathbf{q}_0=\big (q^{(6)},q^{(4)}_{I},q^{(2)}_{I},q^{(0)}\big )^{\rm T}\ , \qquad I =1, \ldots, h^{1,1}(Y_3)\ , 
\end{equation}
where we indicated with the superscript that~$q^{(p)}$ will later be interpreted as inducing D$p$-brane charges. 
Now we enforce the conditions~\eqref{masslessness} and~\eqref{masslessnessII} by using the explicit forms of the log-monodromies~$N_{(i)}$ given in~\eqref{Nn}. 
The details of the computations can be found in appendix~\ref{app:orbits}.
One immediately finds that one needs to demand 
\begin{align}
q^{(6)} &= 0  \qquad \text{for all} \quad  \type{n} \\[-1mm]
\intertext{and}
q^{(4)}_I &=0  \qquad \text{for} \quad \type{n} = \rIII \ttext{or} \rIV\,.
\end{align}
This last condition can also be set for a type II singularity while still satisfying the infiniteness condition \eqref{cond_orbit}. Similarly $q^{(0)}$ is not constrained by the masslessness conditions~\eqref{masslessness} and~\eqref{masslessnessII} but since it plays no role in \eqref{cond_orbit} it can safely be set to zero. That is, we can choose 
\begin{align}
q_I^{(4)} & = 0  \qquad \text{for} \quad \type{n} = \rII\\
q^{(0)} & = 0  \qquad \text{for all} \quad  \type{n}
\end{align}
So we see that the non-trivial sector of these conditions is for~$q_I^{(2)}$, already hinting that the infinite orbit will correspond to D2-brane states.
The masslessness conditions~\eqref{masslessness} and~\eqref{masslessnessII} are then satisfied, for all singularity types, if \vspace{2mm}
\begin{align}
 \label{omega}
q_I^\two &= \ \KK{IJ}{n} \o^J \\[-2mm]
\intertext{for some integer vector $\o^I$ such that \vspace{-2mm}}
q_i^\two &= 0 \qqtext{for} i < n_\rIII \,, \label{massless1}\\[-2mm]
\intertext{where $n_\rIII$ labels the first type III singularity, and \vspace*{-2mm}}
\sum_{a =1}^n q_a^\two = \KK{I}{n} \o^I &= 0  \qqtext{if} \type{n} = \rIII \,. \label{masslessbis}
\intertext{This last condition is always satisfied if we extend \eqref{massless1} to}
q_i^\two &= 0 \qqtext{for} i < n_\rIV \,.\label{nIV} \\
\intertext{which we will take for simplicity. 
Notice that this latter condition is not necessary unlike  \eqref{massless1} and \eqref{masslessbis}. However, we will see in the appendix that it is possible to find an infinite charge orbit satisfying  \eqref{nIV}. It would be interesting, though, to investigate what changes if it is relaxed; but we leave
this task for future work. \vspace*{2mm} \newline 	
\indent On the other hand, the condition \eqref{cond_orbit} for the orbit to be generated requires that \vspace*{-2mm}}
q_{J^*}^\two & \ne 0 \qquad \text{for some} \quad J^*. \label{infinite}
\end{align}

\vspace*{-3mm}
We outline in appendix~\ref{app:orbits} a concrete approach to find some~$\o^I$ such that eqs.~\eqref{massless1} to \eqref{infinite} are satisfied, ensuring that there always exists a massless infinite charge orbit. It is expected that this can be always achieved, since the existence of an orbit was already shown in~\cite{Grimm:2018cpv} in a more abstract way.
Having determined~$\mathbf{q}_0$ we can derive the charge orbit by acting with the log-monodromies~$N_I$ as in~\eqref{Q-N}. This yields  
\begin{align} \label{general_Q}
\mathbf Q  = \Big( 0, 0,\ldots,0, q_I^{(2)},-\textstyle \sum_I m_I q_I^{(2)}  \Big)^\T\ ,
\end{align}
where ~$q_I^{(2)}$ meets the above requirements.

In Type II compactifications, this orbit of states has a specific microscopic interpretation in terms of BPS wrapping D-brane states. For concreteness, in a Type IIB compactification on a Calabi-Yau threefold~$\tilde Y_3$,~$\mathbf{q}$ would correspond to the charge of a D3-brane wrapped on the three-cycles~$\gamma^\cI$ and whose mass~$M=|Z(\mathbf{q})|$ would be given by the central charge
\begin{equation} \label{def-Z}
Z(\mathbf{q}) = e^{\frac{K_{\rm cs}}{2}} \int_{Y_3} H \wedge \Omega = \frac{ \mathbf{\Pi}^{\rm T}\,\pairing\,\mathbf{q}}{\big(i\mathbf{\bar \Pi}^T \pairing \mathbf{\Pi})^{1/2}}\ .
\end{equation}
Here,~$H$ is the three-form with coefficients~$\mathbf{q}$ in the integral basis~$\gamma_I$ and the periods~$\mathbf{\Pi}$ and the K\"ahler potential~$K_{\rm cs}$ in the complex structure moduli space are defined in~\eqref{Kcs} and~\eqref{Pics}. The masslessness conditions~\eqref{masslessness} and~\eqref{masslessnessII} are obtained from requiring that~$Z(\mathbf{Q})=0$ at the infinite distance singularity. 

By using the mirror map, it is also possible to translate these results to the K\"ahler moduli space of Type IIA Calabi-Yau compactifications. The D3-branes will map to different bound states of D$p$-branes with even~$p$. More precisely, notice that we have conveniently chosen a basis for the mirror period vector in~\eqref{tI-periods}, which is identified with the following Type IIA K-theory basis of branes,
\begin{equation} \label{basis_cO}
(\cO_{Y_3}, \cO_{D_I}, \mathcal{C}^J, \cO_p),
\end{equation}
where~$p$ are points,~$D_J$ are~$h^{1,1}(Y_3)$ divisors and~$\mathcal{C}^J := \iota_!\cO_{C^J}\big(K^{1/2}_{C^J}\big)$ where~$C^I$ are the dual~$h^{1,1}( Y_3)$ curves, so~$C^J \cdot D_I =\delta_I^J$ (see~\cite{Gerhardus2016}, section 2.3 for their precise definition). Recall that the divisors~$D_I$ are Poincar\'e-dual to the two forms~$\omega_I$ in~\eqref{t_IIA} and span the K\"ahler cone. In practice, this implies that the different components of the charge vector~$\mathbf{q}$ correspond to the charge of a D6-,D4-,D2- and D0-brane wrapping the whole threefold~$Y_3$, a 4-cycle, a 2-cycle or a point respectively. Therefore, the massless infinite charge orbit at large volume consists of D2-D0 bound states.

It might seem surprising that we are identifying the massless tower predicted by the Swampland Distance Conjecture at the large volume limit of Type IIA with a massless charge orbit of BPS states consisting of bound states of D-branes instead of Kaluza-Klein states. Clearly, there can be more than an infinite tower becoming massless at infinite distance as we will also get a KK tower in this limit. However, it is this charge orbit of BPS states the one that will be later identified as responsible for emergence of the infinite distance and restoration of a global symmetry. Let us also remark that these BPS states only become massless with respect to the Planck scale~$\mp$, since the central charge gives the value of the mass in Planck units. Since the Planck mass is also going to infinity in the large volume limit, the states become indeed infinitely heavy but their mass diverges exponentially slower than~$\mp$. The massless requirement of the Swampland Distance Conjecture only makes sense then in the Einstein frame, where~$\mp$ is kept finite.

\subsection{Infinite distances and charge orbits in elliptic fibrations\label{sec:infinite_elliptic}}

In this section we will determine the singularity types and charge orbits arising in elliptic fibrations with a single section. This analysis will be very useful in the context of the M/F-theory duality in section~\ref{MF-section}. 
In order to do that one first needs to determine the K\"ahler cone basis for these 
geometries. This was done, for example, in ref.~\cite{Cota:2017aal}. 

We denote the base of this elliptic fibration by~$B_2$ and introduce the map~$\pi: Y_3 \rightarrow B_2$ projecting onto~$B_2$. We will assume that~$B_2$ admits a simplicial K\"ahler cone basis, which we then pull to two-forms~$\omega_\alpha$ on~$Y_3$ via~$\pi^*$.
On the threefold~$Y_3$ the two-form cohomology naturally splits as 
\begin{align}\label{tilde_basis}
\tilde \o_I = \{\tilde \o_0, \o_\a\}\ ,
\end{align} 
where~$\tilde \o_0$ is Poincar\'e-dual to the base divisor~$B_2$ and  the~$\o_\a$ are Poincar\'e-dual to divisors~$D_\a = \pi^{-1}(D_\a^{\rm b})$, which are inherited from divisors~$D_\a^{\rm b}$ in the base. This amounts to say that 
$h^{1,1} (Y_3) = h^{1,1} (B_2) +1$.
One can show that the intersections numbers~\eqref{intnbIIA} are then given by 
\begin{equation}\label{int_numb}
\begin{alignedat}{3}
& \tilde{ \c K}_{000} = \base_{\a\b} K^\a K^\b \, , \qquad \quad && 
\tilde{ \c K}_{00\a} && =   \base_{\a\b} K^\b \ ,\\
& \tilde{ \c K}_{0\a\b}  = \base_{\a\b} \, , &&  
\tilde{ \c K}_{\a\b\gamma} && = 0\ .
\end{alignedat}
\end{equation}
where~$\base_{\a\b} = D_{\a}^{\rm b} \cdot  D_{\b}^{\rm b} = B_2 \cdot  D_\a \cdot D_\b$ is a non-degenerate symmetric matrix with signature~$(1,h^{1,1}(B_2)-1)$ and~$K^\alpha$ are the expansion coefficients of the first Chern class of the base~$c_1(B_2) =- K^\alpha \omega_\alpha$.\footnote{Note that in this expansion one actually has to use the two-forms on~$B_2$, but we abuse notation slightly.}
In order to obtain a K\"ahler cone generator in the~$\tilde \omega_0$ direction one has to perform the shift
\begin{equation}\label{Kcone}
\omega_0 = \tilde \omega_0 - K^\alpha \omega_\alpha\ ,
\end{equation}
This implies that intersection numbers in the K\"ahler cone basis~$\omega_I = \{\omega_0,\omega_\a\}$ are given by 
\begin{equation}\label{int_numb_Kc}
\begin{alignedat}{3}
& \c K_{000} = \base_{\a\b} K^\a K^\b \, , \qquad \quad  && \c K_{00\a} && =   - \base_{\a\b} K^\b,\\
& \c K_{0\a\b}  = \base_{\a\b} \, ,   \qquad && \c K_{\a\b\gamma} && = 0\ .
\end{alignedat}
\end{equation}
We note that all these intersection numbers are positive, as required in the K\"ahler cone, for~$h^{1,1}(B_2)\leq 10$,
since also~$\int_{B_2} \omega_\a \wedge c_1(B_2) = - \base_{\a\b} K^\b
\geq0$. The K\"ahler form 
can be also expanded in this basis 
\begin{equation}
J = v^I \omega_I= v^\alpha \omega_\alpha +  v^0 \omega_0\ ,
\end{equation}
which defines the cone~$v^0,v^\alpha > 0$.

Using these intersection numbers and the rules in Tables~\ref{Type_Table} and~\ref{Type_Table2} 
we can read off the singularity types
if some or all of the~$h^{1,1}(Y_3)$ coordinates are taken into a limit. 
Since~$\K{\a\b\gamma}=0$, the only way to obtain a Type IV singularity is to send~$v^0 \to \infty$. Considering first that situation, we find that there are only two cases, depending on whether~$v^0$ is the only coordinate taken to infinity or not. In the first case, the singularity is of Type IV$_{\hoo{B_2}}$, while in the second case we find a singularity of Type IV$_{\hoo{Y_3}}$, which is the maximal singularity type, already when a single coordinate is added to the limit. That is, we have 
\begin{subequations}\label{v0}
	\begin{align}
	v^0  & \rightarrow \infty  : \qquad \text{Type IV}_{h^{1,1}(B_2)}\ ,\label{singI}\\
	v^0, v^1, \ldots, v^n & \rightarrow  \infty  : \qquad \text{Type IV}_{h^{1,1}(Y_3)}\ ,\label{singII}
	\end{align}
\end{subequations}
where in the second limit, the number~$n$ of coordinates~$v^\a$ is non-zero but otherwise arbitrary. The second situation is when~$v^0$ stays finite, i.e. we take the limit~$v^1, \ldots, v^n \to \infty$ with~$n$ arbitrary.\footnote{Recall that the ordering of the coordinates is also arbitrary, meaning that we do not impose any restriction on which of the~$v^\a$ we choose.} Here again we find two cases, depending on whether all the~$\base_{ij}$ vanish or not:
\begin{subnumcases}{v^1,\ldots, v^n \rightarrow  \infty: \qquad \label{vi}}
\text{Type } \rII_2 & if~$\quad \base_{ij} = 0 \quad \forall \ i,j=1,\ldots,n$ \label{singIII}\\
\text{Type }\rIII_0 & otherwise \label{singIV}
\end{subnumcases}

With this at hand, we find that there are only three possible enhancement chains (of course sub-chains of the last one are also possible)
\begin{align}\label{enhanc}
\begin{split}
&\xrightarrow{ v^0\to \infty} \text{IV}_{\hoo{B_2}} \xrightarrow{v^\a \to \infty} \text{IV}_{\hoo{Y_3}} \,, \\
&\xrightarrow{v^\a\to \infty} \text{II}_{2} \xrightarrow{ v^0 \to \infty} \text{IV}_{\hoo{Y_3}}\,, \\
&\xrightarrow{v^\a\to \infty} \text{II}_{2} \xrightarrow{v^\b\to \infty} \text{III}_{0} \xrightarrow{v^0 \to \infty} \text{IV}_{\hoo{Y_3}} \,,
\end{split}
\end{align}
where the conditions on the~$v^\a$'s for these to happen can easily be read off~\eqref{v0} and~\eqref{vi}.

Having determined the arising singularity types we can use the results of the previous section to obtain the 
charge orbit. As described there, 
this first requires to 
determine the growth sector~\eqref{growth-sector} in which the considered path~$t^I(s)$ towards a point 
$P$ at a limiting point~$t^1,\ldots,t^n \rightarrow i \infty$. This might require to reorder the coordinates, in the 
sense that~\eqref{growth-sector} is only satisfied along a path if we permute the coordinates in~\eqref{growth-sector}.
In elliptic fibrations the crucial information required to determine the orbit is 
the growth of~$v^0$ compared to the~$v^\alpha$'s. 
Let us first assume that we have picked an ordering of the~$v^\alpha$'s
such that the path is in the corresponding growth sector. We then relabel 
these~$v^\alpha$'s, such that the ordering is simply~$\big(v^1,\ldots,v^{h^{1,1}(B_2)} \big)$, 
where we are free to pick any ordering for the coordinates that are not sent into a limit. 
We next ask in between which two elements~$v^{\hat n-1}$ and~$v^{\hat n}$ the 
$v^0$ lies, i.e.~for which~$\hat n$ one has
\begin{equation}\label{v0_growth}
\frac{v^{\hat n-1}}{v^0} > \lambda\ ,  \qquad \frac{v^0}{v^{\hat n}} > \lambda\ . 
\end{equation}
The integer~$\hat n$ determines at which point in the enhancement chain a 
Type IV singularity occurs, as explained above. 
It follows from eq.~\eqref{nIV} that all~$q^{(2)}_1,\ldots,q^{(2)}_{\hat n-1}$ are vanishing, while~$q_0^{(2)}$ is the first possibly non-vanishing charge, if we order the charges according to the order of the coordinates appearing in the growth sector. However, for later convenience,
we will adopt a different ordering, namely that~$q_0^\two$ is always the last of the~$q_I^\two$'s, even though~$v^0$ grows faster than the~$v^i$'s with~$i\geq \hat n$, as indicated above.  This ordering will be useful when discussing the interpretation of the charge orbit in F-theory. 
Using~\eqref{general_Q} with \eqref{nIV}, we find
\begin{equation}\label{orbit_ell}
\mathbf Q 
= \Big( {0,\ldots,0},q_{\hat n}^{(2)},  \ldots, q_{\hoo{B_2}}^{(2)},q_0^{(2)},-m_0q_0^{(2)}-\textstyle \sum^{h^{1,1}(B_2)}_{i = \hat n}m_i q^{(2)}_{i}\Big)^{\rm T}\, ,
\end{equation}
where at least one of the~$q_I^{(2)}$ has to be non-vanishing, as required by eq.~\eqref{infinite}.

Actually we show in appendix~\ref{app:orbits} that it is always possible to choose the~$w^I$ in~\eqref{omega} such that \emph{only}~$q_0^\two$ is non-vanishing. That is to say, for \emph{any} path towards the large volume point, one can find the following massless infinite orbit
\begin{equation}\label{orbit_ell_q0}
\mathbf Q 
= \Big( {0,\ldots,0},q_0^{(2)},-m_0q_0^{(2)}\Big)^{\rm T}\ .
\end{equation}
Furthermore, the presence of this orbit is independent of the intersection numbers, so it is valid for any Calabi-Yau threefold. This is one of the central results of this section and will be especially important in section \ref{sec:Ftheorylimit} when studying the F-theory limit.

Let us close this section by briefly discussing the sector dependence of these results. 
Crucially, as stated in~\eqref{v0_growth}, the form of the charge orbit~\eqref{orbit_ell} in general depends on the growth of~$v^0$ relative to the~$v^\alpha$. However, it is also immediate from the occurring singularities listed~\eqref{v0} and~\eqref{vi}
that the relative growth of the~$v^\a$,~$\a \leq \hat n-1$ and~$v^\a$,~$\a \geq \hat n$ is irrelevant to the form of~$\mathbf{Q}$. Hence, we find that for elliptic fibrations the large volume 
charge orbit~\eqref{orbit_ell} exhibits a much milder path-dependence than what generally arises due to the presence of growth sectors. 
In particular, the special choice of orbit~\eqref{orbit_ell_q0} is completely independent of the path.

\subsection{Transferring the orbit to small volumes\label{sec:smallfiber}}

In the previous subsections we have discussed the charge orbits arising in the large
volume regime. In particular, we have generally constructed an infinite orbit~$\mathbf{Q}$ in~\eqref{general_Q}
that becomes massless at a point~$P$ in the large volume regime. 
We might now ask if we can carry this orbit to other points in moduli space away from large volume. 
In general, this is an extremely hard question, since it requires information about the global properties of the 
moduli space and the D-brane states existent at various other points. For elliptic fibrations, however, there
is much literature~\cite{Candelas:1994hw,Klemm2012,Alim:2012ss,Huang:2015sta,Haghighat:2015qdq,Cota:2017aal} on how to leave the large volume point using the map~$v^0 \rightarrow 1/v^0$, 
where we recall that~$v^0$ is the volume of the elliptic fiber.
In the following, we will use these results to present a charge orbit for the limit 
\begin{equation}
v^0  \equiv \frac{1}{\tilde v^0} \, \rightarrow 0 \ .
\end{equation}
Note that this corresponds to considering a completely different region in moduli space 
as indicated in Figure~\ref{fig_lv_sv}. As a byproduct we thus find an example that there can be infinite massless orbits 
at singularities in moduli space that do not satisfy the conditions outlined in subsection~\ref{Kahler_charge_orbits}. It was shown in~\cite{Garcia-Etxebarria:2014wla} that the monodromy transformation associated to the small fiber divisor can be of finite order if the number of sections of the mirror dual is not high enough. In these cases, the divisor~$v^0\rightarrow 0$ is of type I (finite distance) and the intersection point with large base volume will be at most type III$_0$. Hence, there does not exit any local monodromy operator that can generate a massless infinite charge orbit at the regime of small fiber, but still there should be an infinite massless tower of states since the intersection point with large base volume is always at infinite distance. Interestingly, it turns out 
that we can still identify an infinite charge orbit which is transferred from points that satisfy the conditions of subsection~\ref{Kahler_charge_orbits} as suggested 
in~\cite{Grimm:2018cpv}.  In particular, the orbit is transferred from the large volume point as we explain in the following.

\definecolor{myred}{rgb}{0.7, 0.11, 0.11}
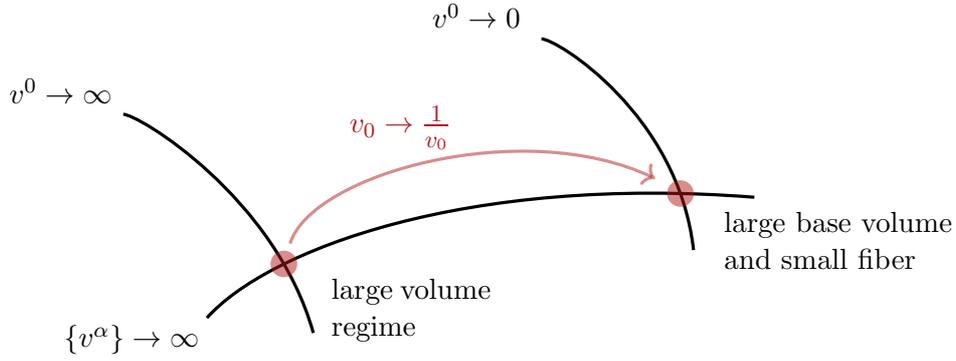
\begin{figure}[ht]
	\begin{center}
		\begin{tikzpicture}
		\draw[name path = va,very thick] (-0.4,-0.2) node [below left=-0.5mm] {$\{v^\a\} \to \infty$} .. controls (0.4,0.7) and (3,1.7) .. (6.8,1.4);
		\draw[name path = v0i,very thick] (-1.5,2.5) node [above left] {$v^0 \to \infty$} .. controls (-1.3,2.5) and (0.5,1.4) .. (1,-0.4);
		\draw[name path = v00,very thick] (2.5+1.5,3.5) node [above left] {$ v^0 \to 0~$} .. controls (2.7+1.5,3.5) and (5.8,2.5) .. (6,0.7);
		\fill[name intersections={of=va and v0i},very thick, myred,opacity=0.5]  (intersection-1) circle (5pt) node [black,below right =0.5mm,opacity=1] { \hspace*{3mm} \parbox{2.1cm}{large volume regime}};
		\fill[name intersections={of=va and v00},very thick,myred,opacity=0.5]  (intersection-1) circle (5pt) node [black,below right=1mm,opacity=1] { \hspace*{2mm} \parbox{3cm}{large base volume and small fiber}};
		\draw [name intersections={of=va and v0i, by=x}][name intersections={of=va and v00, by=y}]
		[very thick,myred,opacity=0.5,->,shorten >=11pt,shorten <=8pt] (x) .. controls (1,1.8) and (4.2,2.6) .. (y) node [pos=0.4, above =1.5mm,opacity=1] {$v_0 \to \mmfrac{1}{v_0}$} ;
		\end{tikzpicture}
	\end{center}
	\caption{The large volume regime is related to the  small fiber regime by a double T-duality along the elliptic fiber. This duality 
		is implemented by a Fourier-Mukai transform.}
	\label{fig_lv_sv}
\end{figure}

Considering first Type IIA string theory on a two-torus of volume~$v^0$, it is well-known that the map~$v^0 \rightarrow 1/v^0$
arises from applying T-duality along both torus circles. The basic idea is to apply this to elliptic fibrations 
by performing the double T-duality along the fiber.
To implement this transformation one performs a so-called Fourier-Mukai transformation. This transformation 
acts as a non-trivial linear map~$S$ acting on the K-theory basis of D-branes
\begin{equation} \label{K-basis-ellf}
(\cO_{Y_3},\cO_{D_0},\cO_{D_\alpha},\mathcal{C}^\alpha,\mathcal{C}^0,\cO_{pt})\ , 
\end{equation} 
which is the specialization of~\eqref{basis_cO} to elliptic fibrations.
The  form of the matrix~$S$ can be explicitly calculated following~\cite{andreas2000fourier,andreas2004fourier,Cota:2017aal} as we 
show in detail in appendix~\ref{FM_appendix}. The resulting expression acting on the basis~\eqref{K-basis-ellf} takes the form 
\begin{equation}\label{S-explicit}
S=\left(\ \ 
\begin{array}{cccccc}
\cline{1-2}
\multicolumn{1}{|c}{0}&\multicolumn{1}{c|}{1}&0&0&0&0\\
\multicolumn{1}{|c}{-1}&\multicolumn{1}{c|}{0}&0&0&0&0\\\cline{1-2}
K^\a & 0 & 0 & \base ^{\a \b } & 0 & 0 \\
-K_\a & -K_\a & -\base _{\a \b} & 0 & 0 & 0 \\ \cline{5-6}
0 & 0 & 0 & \frac 12 \left(K^\b-\base^{\b\gamma} \base_{\gamma\gamma}\right) & \multicolumn{1}{|c}{0} &  \multicolumn{1}{c|}{1} \\
0 & \frac 12 K^\gamma \left(\base_{\gamma\gamma}- K_\gamma \right) & \frac 12 \left(\base_{\b \b} -K_\b\right) & 0 &  \multicolumn{1}{|c}{-1}  &  \multicolumn{1}{c|}{0}  \\
\cline{5-6}
\end{array}
\ \ \right) \ ,
\end{equation}
where~$K_\alpha = \base_{\alpha \beta } K^\b$.
One checks that this transformation preserves the symplectic inner product~$\pairing$ given in~\eqref{eta-lcs}, i.e.~that~$S^T \pairing S =\pairing$. 
Note that~$S$ contains, as indicated with the boxes, 
the standard S-duality matrix. As we will see momentarily this is in accord with the fact that 
the double T-duality along the fiber maps
$t^0 \rightarrow -1/t^0$, which is the non-linear S-duality transformation of the complex parameter~$t^0$.
Furthermore, we also stress that~$S$ transforms the D-brane states supported in the elliptically fibered 
geometry. Recalling it corresponds to a double T-duality on the elliptic fiber we find, in particular that
\begin{equation}
\left( \begin{array}{c} \text{D2}_{\rm f} \\ \text{D0} \end{array}\right) \ \xrightarrow{S} \     \left(  \begin{array}{c}  \text{D0} \\
\text{D2}_{\rm f} \end{array} \right)\ ,
\end{equation}
where~$\text{D2}_{\rm f}$ are the D2-branes wrapped on the elliptic fiber. 

This duality operation also relates the periods~$\mathbf{\Pi}$ valid at the large~$v^0$ regime to the small~$v^0$ regime. 
In particular, it relates the large volume central charges as 
\begin{equation}\label{central_rot}
\Big| Z \big[ S \mathbf{\Pi}(  t^\alpha, t^0)\big] \Big| = \bigg| Z \Big[  \mathbf{\Pi} \Big (   t^\alpha + \mmfrac 12 \, k^\alpha, - \mmfrac{1}{t^0} \Big)\Big] \bigg| \ ,   
\end{equation}
where~$S$ is the matrix given in~\eqref{S-explicit}. This expression means that one can equate the 
central charges~\eqref{central_rot} when either replacing the periods~$\mathbf{\Pi} \rightarrow S\mathbf{\Pi}$ 
or evaluating the periods at a different coordinate location. 
Note that if the left-hand side are the large volume periods valid at  
$v^0 = \I \, t^0 \gg 1$ and~$v^\a =\I\, t^\alpha \gg 1$
the right-hand side is now valid in the 
regime~$\I\, \tilde t^0 = 1/ v^0 \gg 1$ and~$v^\alpha \gg 1$. 
It is non-trivial to show~\eqref{central_rot}, since it equates central charges at different points 
in moduli space. However, it was argued in~\cite{Candelas:1994hw,Klemm2012,Alim:2012ss,Huang:2015sta,Haghighat:2015qdq,Cota:2017aal} that the transformation~$S$ effectively 
maps 
\begin{equation}\label{t->1/t}
t^0 \ \mapsto - \mmfrac{1}{t^0}\ , \qquad \tilde t^\alpha \ \mapsto \tilde t^\alpha + \mmfrac 12 \, k^\alpha \ , 
\end{equation}
when explicitly evaluating the power series expansions of the periods. An arising overall 
complex rescaling of~$\mathbf{\Pi}$ can be absorbed by a transformation of the K\"ahler potential appearing in 
the central charge~$Z$ leading to~\eqref{central_rot}.

In the previous section we gave in~\eqref{orbit_ell} the massless infinite 
charge orbit at the large volume point for an elliptic fibration. In particular, 
this orbit is at large fiber volume,~$v^0 \to \infty$.
In order to obtain the orbit at small fiber volume we note that~\eqref{central_rot}
implies that if~$\mathbf{Q}_{\rm LV}$ is the large volume orbit massless at 
$t^0,t^1,\ldots,t^n \rightarrow i \infty$, the orbit 
\begin{equation} \label{QF=SQ}
\mathbf  Q_{\rm F}  = S \, \mathbf Q_{\rm LV} 
\end{equation}
will be massless at~$v^0 \rightarrow 0$. 
Using the explicit expressions~\eqref{S-explicit} and~\eqref{orbit_ell} we find
\begin{equation} \label{QF-sec35}
\mathbf Q_{\rm F}
= \Big( 0,\ 0,\ \base^{\alpha i} q_i^{(2)},0,\ -m_0 q_0^{(2)}-\textstyle \sum_{i} m_i q^{(2)}_{i}+\frac 12 (K^i-\base^{i\a}K_{\a\a 0})q_i^{(2)},\ q_0^{(2)}\Big)^{\rm T}\,,
\end{equation}
where we recall that~$i \ge \hat n$ designates the~$v^\a$ that grow slower than~$v^0$ when taking the limit, see~\eqref{v0_growth}.
In order to read the actual charge, we need to further contract with~$\pairing$  
\begin{equation}
\mathbf Q_{\rm F} \cdot  \pairing
= \Big( -K^i q_i^{(2)}+q_0^{(2)},\ -m_0 q_0^{(2)}-\textstyle \sum_{i} m_i q^{(2)}_{i},\ 0,\ \base^{\a i} q_i^{(2)},\ 0,\ 0\Big)^{\rm T}\ .
\end{equation}

Hence, the infinite tower of states becoming  massless at small volume of the fiber consists of D2-D0 bound states which differ by the D2-brane charge along the elliptic fiber. The orbit can also admit a $D4$-charge although, as remarked in the previous section, is always possible to choose an infinite orbit in which this D4-charge vanishes. The transfer of the orbit from the large volume regime to small fiber is highly non-trivial and highlights the intricate global structure which is required to satisfy the Swampland Distance Conjecture at any infinite distance point of the moduli space.

\section{On infinite distances and charge orbits in M- and F-theory} \label{MF-section}

In this section we will consider M-theory compactified on an elliptically fibered Calabi-Yau threefold~$Y_3$
and the duality of this setting to F-theory on the same threefold~$Y_3$ times an additional circle~$S^1$. We will study infinite distances and charge orbits arising near the large volume point of such an elliptically fibered geometry in M-theory. 
Subsequently we generalize the discussion to include the F-theory limit which requires sending the volume of the elliptic fiber to zero. In the F-theory dual picture this limit corresponds to sending the radius of the additional~$S^1$ to infinity. The resulting 
effective action then describes F-theory compactified on the elliptically fibered~$Y_3$. 
This leads us to a dual geometric realization of the infinite tower of Kaluza-Klein states associated to~$S^1$ in terms of an infinite charge orbit by using the discrete symmetries associated to the large volume regime in M-theory. These discrete symmetries are captured by monodromy transformations when considering the complexified K\"ahler moduli space. 

\subsection{6D Supergravity circle compactification and F-theory} \label{sec:6Dcircle}

In this subsection we review the circle compactification of the 6D~$\cN=(1,0)$ supergravity effective theory obtained from compactifying F-theory on a Calabi-Yau threefold. We first revisit the classical reduction and then include one-loop corrections from the Kaluza-Klein tower.
Our presentation will closely follow~\cite{Bonetti:2011mw}, but we refer to~\cite{Ferrara:1996wv} for an earlier study of this setting.

In a generic 6D supergravity with (1,0) supersymmetry (8 supercharges), 
one can have four type of multiplets (restricting to spin less or equal to two): 
the gravity multiplet, vector multiplets, tensors multiplets and hypermultiplets. 
In order to simplify the discussion, we will consider a theory that has no vector multiplets and 
contains in addition to a 
gravity multiplet~$n_T$ tensor multiplets 
as well as~$n_H$ neutral hypermultiplets. 
To ensure cancellation of gravitational anomalies we will set~$n_H= 273 -29 \,n_T$. 
Note that this limits the number of tensor multiplets that one can consider, as it requires~$n_T \leq 9$.

The bosonic field content of the theory under consideration consists of the graviton~$\hat g_{\mu \nu}$,~$n_T+4n_H$ real scalars, one self-dual and~$n_T$ anti-self-dual two-forms 
collectively denoted by~$\hat B^\alpha$,~$\a=1,\ldots,n_T+1$, whose field-strengths~$\hat G^\a = \d \hat B^\a + \frac 12 a^\a \hat \omega_{\rm grav}$ contain the gravitational Chern-Simons form (see e.g.~~\cite{Bonetti:2011mw} for further details).
The bosonic part of the 6D supergravity (pseudo-) action takes the form 
\begin{equation}\label{S6}
\begin{split}
S_6 = \mpl{6 }^4\int_{\c M_6} \mmfrac 12 \, \hat {\mathbf R } \, \hat \star \,1- \mmfrac 14 g_{\a \b} \hat G^\a \wsh \,\hat G^\b - \mmfrac 12 g_{\a\b} \d j^\a \wsh \d j^\b -  h_{uv} \d \hat q^u \wsh \, \d \hat q^v\qquad\\[-0.3em]
- \mmfrac 14 \, \Omega_{\a\b} \, a^\a \hat B^\b \wedge {\rm Tr} \, \big(\hat{\c R} \wedge \hat{\c R}\big) \ ,
\end{split}
\end{equation}
where the~$\hat q^u,\ u=1,\ldots, 4n_H$ are the scalars in the hypermultiplets. The~$n_T +1$ real scalars~$j^\a$ are subject to the constraint 
\begin{equation}\label{6Dconstr}
\Omega_{\a\b} j^\a j^\b =1,
\end{equation}
where~$\Omega_{\a\b}$ is a constant~$SO(1,n_T)$ metric, leaving effectively~$n_T$ independent real scalars
that reside in the tensor multiplets.
The positive definite, and non-constant, metric~$g_{\a\b}$ of scalar manifold is defined as 
\begin{equation}
g_{\a\b} = 2 j_\a j_\b - \Omega_{\a\b}; \qquad j_\a = \Omega_{\a\b} j^\b.
\end{equation}
The (anti)-self-duality conditions for the two forms~$\hat B^\a$ in a~$SO(1,n_T)$ takes the form 
$g_{\a\b} \, \hat \star \, \hat{G}^\b = \Omega_{\a\b} \, \hat G ^\b$ and has to be 
imposed by hand in addition to the equations of motion derived from the 
action~\eqref{S6}.
Let us note that there is a convenient way to introduce the 
coordinates~$j^\alpha$, such that~\eqref{6Dconstr} is automatically satisfied. 
More precisely, we can introduce real unconstraint scalars~$v_{\rm b}^\a$ and 
define 
\begin{equation}\label{j^a}
j^\a = \frac{v^\a_{\rm b}}{\cV_{\rm b}^{1/2}} \ , \qquad  \cV_{\rm b} = \Omega_{\alpha \beta} \, v^\a_{\rm b}v^\b_{\rm b} \ . 
\end{equation}
Since the~$v^\a_{\rm b}$ are unconstraint there is an extra degree of freedom~$\cV_{\rm b}$. It turns out that 
in F-theory compactifications it is actually physical and resides in a hypermultiplet as we discuss below. 

We now proceed to reduce action~\eqref{S6} on a circle, focusing on the two-derivative part. The 6D metric and 
two-forms~$\hat B^\a$ are reduced as 
\begin{align}\label{AnsatzB}
\d \hat s^2 = \d s ^2 - r^2 (\d y - A^0)^2\ , \qquad \hat B^\a = B^\a - A^\a (\d y - A^0)\ ,
\end{align}
where~$A^0$ is the Kaluza-Klein vector and~$B^\a$ and~$A^\a$ are 5D two-forms and one-forms, respectively. 
Dimensionally reducing the (anti)-self-duality condition to~$r \, g_{\a\b} \star G^\b = -\Omega_{\a\b} F^\b$
we can eliminate the two-forms from the 5D action and only retain 5D vectors.
The five-dimensional Einstein frame action at the two derivative level then takes the form \footnote{This requires a 
	Weyl rescaling of the metric~$g^E_{\mu \nu} = (r/r_0)^{2/3} g_{\mu \nu}$.}
\begin{equation}\label{S5}
\begin{split}
S_5 = r_0 \mpl{6}^4\int_{\c M_{5}} \mmfrac 12\, \mathbf R \star 1 -  h_{uv} \d  q^u \ws \, \d q^v - \mmfrac 23 \, r^{-2} \, \d r \ws \d r - \mmfrac 14 \, r^{8/3} r_0^{-2/3} F^0 \ws F^0 \hspace*{6mm}\\
- \mmfrac 12 \, g_{\a\b} \left( \d j^a \ws \d j^\b + r^{-4/3} r_0^{-2/3} F^\a \ws F^\b \right)- \mmfrac 12 r_0^{-1}\Omega_ {\a\b} A^0 \wedge F^\a \wedge F^\b.
\end{split}
\end{equation}

Since such a circle reduction does not break any supersymmetry this is
a 5D~$\cN=2$ supergravity theory (8 supercharges), 
with one gravity multiplet and 
$\nvf = n_T+1$ vector multiplets, and~$n_H$ neutral hypermultiplets.  
The bosonic field content of such a theory is one graviton,~$\nvf +1$  vectors\footnote{The~$+1$ comes from the gravity multiplet, which contains a vector.} and~$\nvf + 4 n_H$ real scalars. The canonical form of the action is given by
\begin{equation}\label{S5can}
\begin{split}
S_5^{\rm can} = \mpl{5}^3 \int_{\c M_{5}} \mmfrac 12 \, \mathbf R \star 1 -  h_{uv} \d  q^u \ws \, \d q^v \hspace*{7cm}\\
-\mmfrac 12 G_{IJ}\left( \d M^I \ws \d M^J  + \bar F^I \ws \bar F^J\right)  - \mmfrac{1}{12} C_{IJK} \bar A^I \wedge \bar F^J \wedge \bar F^K \, ,
\end{split}
\end{equation}
where all the vectors are denoted collectively as~$\bar A^I, I=0,\ldots, \nvf$, and the~$\nvf+1$ reals scalars~$M^I$ are subject to the so-called very special geometry constraint 
\begin{align} \label{very_spec_constr}
\c N \equiv \frac{1}{3!} C_{IJK} M^I M^J M^K \overset{!}{=} 1,
\end{align}
leaving effectively~$\nvf$ reals scalar degrees of freedom. This cubic potential~$\c N$ specifies entirely the theory at the two derivatives level, the field metric (which coincide with the gauge coupling function) and the Chern-Simons coefficients being given by
\begin{align}\label{gcf}
G_{IJ} = \left[-\mmfrac 12 \partial_I \partial _J \log \c N\right]_{\c N =1} \ ,
\qquad C_{IJK} =  \partial_I \partial _J \partial_K \c N.
\end{align}
Also at the four-derivative level a 5D~$\cN=2$ action is known that includes 
the term arising from the reduction of the last term in~\eqref{S6}. Concretely, the 
5D action with four-derivative terms presented in~\cite{Hanaki:2006pj} includes the term 
\begin{equation}\label{S5RR}
S_5^{\rm grav}= -\frac{1}{4}    \int_{\c M_{5}} c_I\,  \bar A^I \wedge \tr{\c R \wedge \c R} \ . 
\end{equation}

The general action~\eqref{S5can} matches with the action obtained~\eqref{S5} by 
dimensional reduction if we identify the~$n_T+1$ vector multiplets 
$(M^I,\bar A^I)$ as
\begin{subequations} \label{very_spec_coord}
	\begin{alignat}{2}
	M^0 &= r^{-4/3} \,, \qquad &\bar A^0 &= r_0^{-1/3} A^0 \,,\\
	M^\a  &= r^{2/3} j^\a \,, \qquad &\bar A^\a &= r_0^{-1/3} A^\a\, , 
	\end{alignat}
\end{subequations}
together with a cubic potential given by
\begin{align}\label{N}
\c N^F_{\rm class} = \Omega_{\a\b} M^0 M^\a M^\b\,,
\end{align}
and finally the 5D Planck mass is related to the 6D one as in~\eqref{Mp_circle}, i.e.~$\mpl{5}^3 = r_0 \mpl{6}^4$.
Using the definitions~\eqref{very_spec_coord}, one directly finds~$\c N^F_{\rm class} = \Omega_{\a\b} j^\a j^\b$, such that the constraint~\eqref{6Dconstr} indeed implies~$\c N =1$. Also straightforward to get from the~$\c N^F_{\rm class}$ defined above is that the only non-zero Chern-Simons coefficient
in~\eqref{S5can} is~$C_{0\a\b} = 2\, \Omega_{\a\b}$.
Finally, reducing the higher curvature term in~\eqref{S6} and comparing with~\eqref{S5RR} leads to the identification 
$c_\alpha  = -12\, \Omega_{\a \b} a^\b$, 
with~$c_0$ vanishing in the classical reduction.

The action~\eqref{S5can} evaluated with~\eqref{N} includes only the zero modes of the circle reduction. Higher order massive KK modes have not been written down, however they do run in the loops and might generate quantum corrections, 
as we saw in section~\ref{sec:KK}.
In this work we are interested by the quantum corrections to the moduli space metric. However, because of the very special geometry, the field metric~$G_{IJ}$ is related to the Chern-Simons coefficients~$C_{IJK}$ through the cubic potential~$\c N$, such that all the information is already encoded in the Chern-Simons coefficients. In the case of interest, only~$C_{000}$ is being generated by loop corrections, because the KK modes
are only electrically charged under~$A^0$ and not under~$A^\a$. 
Furthermore supersymmetry tells us that there are no further loop corrections beyond one-loop.
The KK-modes contributing to~$C_{000}$ are massive spin-1/2, massive spin-3/2 and massive two-form fields. The computation of this one-loop correction was carried out~\cite{Bonetti:2013ela} and shown to yield the contribution
\begin{align}\label{C000}
C_{000}^\oneloop =\frac{9-n_T}{4}\ .
\end{align}
Such a Chern-Simons term leads to a piece in the cubic potential~$\c N^\oneloop =\frac 16 C^\oneloop_{000} (M^0)^3$ which in turn gives a one loop correction to the field metric~$G^\oneloop_{00} \sim 1/{(M^0)^2}$. 
This contribution alone already induces an infinite distance singularity at~$M^0 \rightarrow \infty$, which is thus generated at one-loop level. 
Adding the classical result~\eqref{N} and the the one-loop result~\eqref{C000}, we find the following total cubic potential for a circle reduction of a 6D theory with~$n_T$ tensor multiplets and without vector multiplets
\begin{align}\label{Ntot}
\c N_{\rm tot}^F = \Omega_{\a\b} M^0 M^\a M^\b + \frac{9-n_T}{24} \, (M^0)^3
\end{align}
We will now discuss how this result is arising in the dual M-theory compactification on 
an elliptically fibered Calabi-Yau threefold.

\subsection{M-theory on a Calabi-Yau threefold and the F-theory match}
\label{sec:Mthy}

Having discussed the dimensionally reduced 
a 6D~$(1,0)$ supergravity action arising from F-theory on a circle, we now 
briefly recall the match of the resulting effective action with a reduction of M-theory 
on an elliptically fibered Calabi-Yau threefold. This implements the F-theory to M-theory duality. 
The circle radius will then be part of the 
K\"ahler moduli space such a threefold. 

To begin with, we will briefly summarize the dimensional reduction of eleven-dimensional supergravity on a Calabi-Yau threefold. 
This reduction is well-known, see e.g.~\cite{Cadavid1995}, and we will follow the notation of~\cite{Bonetti:2011mw}. 
Eleven dimensional supergravity contains in addition to the metric also a three-form~$\hat C_3$ as bosonic fields, 
where the hat now indicates eleven-dimensional objects. 
We now reduce this theory on a Calabi-Yau threefold~$Y_3$, i.e. we take~$\c M_{11} = \mathbb R^{1,5} \times Y_3$. 
The massless fluctuations around the background Calabi-Yau metric
correspond to complex structure deformations and K\"ahler structure deformations. The former are 
part of hypermultiplets and not be of relevance in the following. Rather we will 
focus on the K\"ahler structure deformations. These are obtained as in~\eqref{t_IIA} by expanding the K\"ahler form~$J$ along harmonic (1,1)-forms as~$J=  v^I  \omega_I$, where~$I =1, \ldots, h^{1,1} (Y_3)$.
Likewise, we expand the three-form~$\hat C_3$ in the same basis 
\begin{equation} \label{C3}
\hat C_3 = A^I \wedge  \o_I+ \ldots\ , 
\end{equation}
where the~$A^I$ are all the vectors of the 5D theory and the dots 
indicated terms yielding hypermultiplet scalars irrelevant in the following. 
We thus find~$h^{1,1}(Y_3)$ vectors~$A^I$, of which one resides in the 5D gravity 
multiplet and~$\nvf  = h^{1,1} (Y_3)-1$ reside in 5D vector multiplets. 
The~$h^{1,1} (Y_3)$ scalars~$ v^I$ are expected to comprise the 
scalars in the~$\nvf$ vector multiplets. The apparent mismatch in 
their number is resolved by noting that the overall  
volume of the Calabi-Yau threefold~$\cV$ defined in~\eqref{def-cVIIA} actually resides in a 
hypermultiplet. 
Accordingly, to separate the total volume~$\cV$ and the scalars~$L^I$ in the vector multiplets it is natural to define 
\begin{equation}\label{LI-def}
L^I = \frac{ v^I}{\c V^{1/3}}\ .
\end{equation}
These fields indeed parametrize only~$h^{1,1}(Y_3)-1$ degrees of freedom, since they satisfy
\begin{equation}\label{cubicpot}
\c N^M \equiv \mfrac{1}{3!} \, { \c K}_{IJK} L^I L^J L^K =1\ .
\end{equation}
This condition matches the general very-special K\"ahler constraint~\eqref{very_spec_constr}, such that the fields~$L^I$ can be identified with 
the very special coordinates and~$\c N^M$ the cubic potential of the 5D~$\c N =2$ in its canonical form. One checks that this 
potential indeed allows to match the action obtained by dimensional reduction~\cite{Cadavid1995}.

As mentioned above, the volume~$\c V$ is one of the scalars of the hypermultiplets sector, and its kinetic term is 
\begin{equation}\label{cVkinetic-term}
h_{uv} \d  q^u \ws \,  q^v \supset \mmfrac 14 \d \log \c V \ws \d \log \c V\,.
\end{equation}
The rest of the hypermultiplet sector will not be relevant for us, so we will only mention that the number of such multiplets is given by~$n_H = h^{1,2}(Y_3) +1$, the remaining~$4 h^{1,2}(Y_3) +3$ real scalars coming from the expansion of~$\hat C_3$ (dots in~\eqref{C3}) 
and from the complex structure deformations of~$Y_3$.
We refer to e.g.~\cite{Bonetti:2011mw} for the full metric.

Up to this point the Calabi-Yau space used in the dimensional reduction was general. 
In order apply the duality between M-theory and F-theory we have to further restrict  
$Y_3$ to be two-torus or elliptically fibered. This will then allow us 
to match the 5D setting obtained from M-theory with the F-theory setting 
discussed in section~\ref{sec:6Dcircle}. Furthermore, recalling that we have
restricted our considerations to include only no 6D vector multiplets and only neutral hypermultiplets we further demand that~$Y_3$ is a smooth elliptic fibration (i.e. without exceptional divisors resolving singularities of the fiber) with a single section. This is the situation described in section~\ref{sec:infinite_elliptic} and we refer to it for the notation used. 

In the expansion of the K\"ahler form~$J$ we are free to choose a basis of (1,1)-forms and hence  either can use  the basis~\eqref{tilde_basis} or the K\"ahler cone basis~\eqref{Kcone}. We will use the latter in order to easily connect 
to the analysis of singularities in~\ref{sec:infinite_elliptic}, although the former is usually used in the literature, such as in ref.~\cite{Bonetti:2011mw}.
Using the intersection numbers~\eqref{int_numb_Kc} in the cubic potential~\eqref{cubicpot}, we obtain
\begin{equation}\label{cub_pot_M}
\c N^M = \mfrac 12 \, \base_{\a\b} L^0 L^\a L^\b  - \mfrac 12 \, \base_{\a\b} K^\a (L^0)^2 L^\b+   \mfrac 16 \, \base_{\a\b} K^\a K^\b (L^0)^3.
\end{equation}

The M/F theory duality tells us that we should be able to match this result with the one of section~\ref{sec:6Dcircle}. While the first term of~\eqref{cub_pot_M} can be matched with the classical term~\eqref{N}, and the last term can be matched with the loop correction~\eqref{C000}, the term in the middle does not appear for a circle reduction. 
This implies that the proper duality match requires 
to first perform the shift 
\begin{equation}\label{def-checkL}
\check L^\a = L^\a - \mfrac 12 \, K^\a L^0,
\end{equation}
and performing a similar redefinition for the vectors~$A^\a$. This corresponds in the geometry to take yet a different basis for the two-forms, namely~$\check \omega_0 = \o_0 + \frac 12 K^\a \o_\a$. This shift indeed removes the second term in~\eqref{cub_pot_M}. Finally, to make the matching more transparent, we note that~$\int_{Y_3} c_1^2(B_2) = \base_{\a\b} K^\a K^\b = 10 - h^{1,1} (B_2)$ . 
The cubic potential now reads
\begin{equation}
\c N^M =  \frac 12 \base_{\a\b} L^0 \check L^\a \check L^\b + \frac{10 - h^{1,1} (B_2)}{24} \, (L^0)^3 .
\end{equation}
This result is now straightforwardly matched with~\eqref{Ntot} by identifying
\begin{equation}\label{L-match}
L^0 = M^0, \qquad \check L^\a = M^\a, \qquad \base_{\a\b} = 2 \, \Omega_{\a\b}, \qquad \text{and} \qquad h^{1,1} (B_2) = n_T +1.
\end{equation}
It can also be checked that the overall volume~$\cV$ in the M-theory compactification is identified with the volume of the base~
$\cV_{\rm b}= \frac{1}{2} \base_{\a \b} v^\a_{\rm b} v^\b_{\rm b}$ in the 6D hypermultiplet of the F-theory compactification, 
\begin{equation}
\label{volumes}
\cV = \cV_{\rm b} \ . 
\end{equation}
Notice, though, that~$\cV$ is given in in 11D Planck units while~$\cV_{\rm b}$ is given in string units.
Finally we note for completeness that the~$K^\a$ have to be matched with the Green-Schwarz parameters~$a^\a$ present in~\eqref{S6} as discussed in~\cite{Bonetti:2011mw}.

\subsection{Large volume limits in M-theory} \label{lv_Mtheory}

Infinite distance limits in  K\"ahler moduli space of an elliptically fibered Calabi-Yau threefold were studied in section~\ref{sec:infinite_elliptic}. The same classification obtained at large volume applies here for a Calabi-Yau threefold compactification of M-theory. However, the microscopic interpretation of the infinite charge orbits in terms of wrapping branes changes. In this section, we will discuss the M/F-theory interpretation of the infinite massless charge orbits obtained at the different large volume limits.

Even if the monodromy transformation has a more obscure meaning in M-theory (since the 5D moduli space is not complex), it is still a very useful tool to classify the infinite distance limits and the tower of states becoming massless. When further compactifying on a circle, we can complexify the moduli space and connect with the IIA interpretation in which the monodromy transformation corresponds to a discrete shift of the axion partners of the K\"ahler deformations~$v^I$. These axions arise from dimensionally reducing the 5D vector bosons~$A^I$ along the extra circle. Therefore, in the 5D M-theory compactification, the monodromy transformations capture the change on the geometry under large gauge transformations of these vectors~$A^I$. At infinite distance, the axionic discrete shift symmetries in Type IIA enhance to a continuous global symmetry. Analogously, in M-theory the discrete shifts of the gauge bosons also become continuous and we restore a one-form global symmetry at infinite distance. The tower of states of the SDC can, therefore, again be understood as a quantum gravity obstruction to restore this generalized global symmetry. 

For the scope of this section, it is enough to recall that we can borrow the results for the classification on infinite distance singularities and charge orbits of section~\ref{sec:infinite_elliptic}. The only difference is that the infinite charge orbit becoming massless at infinite distance will now consist of M2-brane states wrapping certain 2-cycles of the compactification manifold. Recall that even if their masses generically diverge, they become massless with respect to the Planck scale (which diverges exponentially faster). Notice also that the charge orbits obtained in~\eqref{orbit_ell} imply that the tower consists only of particles coming from wrapping M2-branes and not strings coming from M5-branes, since the M5-brane has to vanish in an orbit that 
satisfies the masslessness conditions~\eqref{masslessness},~\eqref{masslessnessII}.

In the following, we will translate these limits and orbits to the F-theory setup.
We recall that the real scalar fields~$j^\alpha$ in the 6D tensor multiplets together with the circle 
radius~$r$ form the coordinates that are identified with the K\"ahler cone coordinates~$v^0, v^\alpha$  through~\eqref{L-match}, together with~\eqref{very_spec_coord},~\eqref{LI-def}, and~\eqref{def-checkL}. One finds
\begin{equation}\label{rj-va-map}
\frac{v^0}{\c V^{1/3}} = r^{-4/3} \ ,\quad \frac{v^\a- \frac 12 K^\a v^0}{\c V ^{1/3}} = r^{2/3}j^\a\ .
\end{equation}
In addition, we have to consider the volume~$\cV$ of the Calabi-Yau threefold defined in~\eqref{def-cVIIA}, 
which is part of a 5D hypermultiplet. In terms of the K\"ahler cone coordinates it reads
\begin{align} \label{cV-in_tildeva}
\c V = \tfrac{1}{2} \c K_{0 \a\b} v^0 v^\a v^\b +  \tfrac{1}{2} \c K_{0 0 \a} v^0 v^0  v^\a +  \tfrac{1}{6} \c K_{0 00} v^0 v^0 v^0\ .
\end{align}
As mentioned in~\eqref{volumes} this volume has to be identified with the volume~$\cV_{\rm b}$ in 
the 6D hypermultiplet.
To recall the charge orbits we stress that the matching with F-theory should be done in the basis of two-forms 
$\check \omega_I=\{\check\omega_0,\omega_\alpha\}$, as explained in section~\ref{sec:Mthy}. 
This basis is related to the K\"ahler cone basis~$\{\omega_0,\omega_\alpha\}$ via
\begin{equation}
\label{KK2form}
\check\omega_0=\omega_0+\mfrac12 K^\alpha \omega_\alpha \ .
\end{equation}
The charge of the states in the orbit under the 5D vector bosons~$A^I$,~$I=\{0,\alpha\}$, is given by~$q_I=\int_{Y_3} H \wedge \check \omega_I$, 
where~$A^0$ corresponds to the Kaluza-Klein vector of the circle reduction and~$A^\alpha$ arise from dimensionally reducing the 6D tensor gauge fields~$\hat B^\alpha$,~$\alpha=1,\dots, n_T+1$.

We begin our analysis of the limits in F-theory moduli space with the large volume limits, in which one or several~$v^I\rightarrow \infty$. Notice that they always imply~$\cV\rightarrow \infty$
and thus always require to take the limit~$\c V_\bb \to \infty$ in in F-theory. As seen from the kinetic term~\eqref{cVkinetic-term}
this limit in the hypermultiplet sector lies at infinite distance. Therefore, these limits are in general at infinite distance both in the tensor and hypermultiplet sectors. 
In section~\ref{sec:infinite_elliptic} we analyzed such limits for elliptic fibrations and we found that only four possible types of singularities were possible, listed in equations~\eqref{v0} and~\eqref{vi}. Here we will study what these limits 
correspond to in the F-theory moduli space by determining the associated behavior of~$r$ and~$j^\a$. 
For simplicity, we will consider the case that all~$v^{\a_i}$ that are taken to a limit grow at the same rate, but note that the generalization to specific growth sectors is straightforward.
The results are summarized in Table~\ref{sing_Fthy}.
\begin{table}[!ht]
	\begin{center}
		\renewcommand{\arraystretch}{1.4}
		\setlength{\tabcolsep}{3mm}
		\begin{tabular}{>{\kern-2mm}lllllllll<{\kern-2mm}}
			\topline \headcol  
			&\multicolumn{3}{c}{growth of} &&&\multicolumn{3}{c}{growth of}  \\
			\headcol
			Singularity &$v^0$&$v^{\a_i}$ &$v^{\a_p}$ &$\base_{\a_i \a_j}$ & Type &$r$ &$j^{\a_i}$ &$j^{\a_p}$ \\	\midline  
			\eqref{singI} &$\lambda$ & - & - & - &$\rIV_{\hoo{B_2}}$  & -& -& - \\[2pt]
			\rowcolor{gray!7}&&&&$= 0$  & &$\lambda^{1/4}$ &$\lambda^{1/2}$ &$\lambda^{-1/2}$ \\
			\rowcolor{gray!7} 
			\multirow{-2}{2cm}{{\eqref{singII}}}   & \multirow{-2}{5mm}{$\kappa$} & \multirow{-2}{*}{$\kappa_\a$} & \multirow{-2}{*}{-} &$\neq0$ & \multirow{-2}{*}{$\rIV_{\hoo{Y_3}}$}   &$\lambda^{1/2}$ & - &$\lambda^{-1}$  \\[2pt] 
			\eqref{singIII}  & - &$\lambda$ & - &$= 0$ &$\rII_2$&$\lambda^{1/4}$ &$\lambda^{1/2}$ &$\lambda^{-1/2}$ \\[2pt] 
			\rowcolor{gray!7} 
			\eqref{singIV} & - &$\lambda$ & - &$\neq0$ &$\rIII_0$&$\lambda^{1/2}$ & - &$\lambda^{-1}$  \\[2pt]
			\bottomline
		\end{tabular}
		\caption{Large volume singularities in terms of the F-theory coordinates~$r$ and~$j$'s. We collectively denoted~$v^{\a_i}, \ i=1,\ldots,n$ the coordinates that are taken in the limit and~$v^{\a_p}, \ p= n+1,\ldots,\hoo{B_2}$ those that are not. In the second line, we defined~$\lambda = \kappa_\a / \kappa$ and assumed~$\lambda \to \infty$. If~$\lambda\to 0$, the result is the same as the one of the first line.}
		\label{sing_Fthy}
	\end{center}
\end{table}

We stress that the first limit~$v^0 \rightarrow \infty$ in Table~\ref{sing_Fthy} is special, since it  lies at finite distance in the tensor moduli space. 
However, as discussed above, it will be still at infinite distance in the hypermultiplet sector, since~$\cV_{\rm b}\rightarrow \infty$. 
All the other limits in Table~\ref{sing_Fthy} correspond to a large radius limits~$r \to \infty$.
In terms of the volumes of the base, for each volume~$v^{\alpha_i}\rightarrow \infty$ there is also a volume of a two-cycle of the base that grows to infinity. 

Finally, let us briefly comment on the F-theory interpretation of the charge orbits arising in the large volume 
limits in the M-theory. Recall that for Type IIA compactifications we have determined the infinite charge orbits 
that become massless at the singularity in~\eqref{orbit_ell}. Considering either of the two situations displayed in the 
last three lines of Table~\ref{sing_Fthy}, the corresponding Type IIA charge orbit reads
\begin{equation}
\label{Q1}
\mathbf Q = \Big(0, {0,\ldots,0},0,q^{(2)}_{\a_p} ,q^{(2)}_{0}, -m_0 q^{(2)}_{0}
-\sum_{\alpha_p} m_{\a_p} q^{(2)}_{\a_p}\Big)^{\rm T}\ ,
\end{equation}
where we recall that the~$\a_p$ label the directions in the base that are not taken to a limit. 
To lift this result to M-theory we note that D2-D0 bound-states correspond to M2-branes with 
a certain KK-charge around the circle~$\hat S^1$ connecting Type IIA and M-theory. 
The last entry of~\eqref{Q1} corresponds to the D0 charge, we realize that this orbit simply 
represents the KK-tower of an M2-brane state wrapped on the curve~$q^{(2)}_{\a_p} \cC^{\a_p}+q^{(2)}_{0} \cC^0$ in~$Y_3$
with all possible KK-charges along~$\hat S^1$. 
Further following the duality to F-theory the M2-brane state encoded by~\eqref{Q1} maps to 
a particle arising from a 6D string wrapping the F-theory circle~$S^1$ to 5D, since for~$q^{(2)}_{0} \neq 0$ and some~$q^{(2)}_{\a_p} \neq 0$  one finds a charge both under the Kaluza-Klein gauge vector~$A^0$ and the gauge bosons~$A^{\a_p}$  associated to the  base. These strings arise from D3-branes in Type IIB wrapping the non-trivial two cycles~$q^{(2)}_{\a_p} \cC^{\a_p}$ in the base whose volume is not sent to infinity. Let us remark that each tower of particles (one per each $q_I^{(2)}\neq 0$) lifts to a single 6D string. Since the volume of the base goes to infinity, all such strings become exponentially light compared to the Planck scale. This is somewhat analogous to the analysis in \cite{Lee2018} in which a 6D string becomes tensionless in the infinite distance limit of sending the gauge coupling of an open string U(1) to zero. 
Note, however, that the latter limit does not correspond to a decompactification limit of the internal space 
and, in particular, keeps $\cV_{\rm b}$ finite. 
To implement such a limit one has to send some subset of coordinates to infinity, while sending others 
to zero. We will discuss an example of such a mixed limit next. 

\subsection{F-theory limit and geometric realization of the Kaluza-Klein tower\label{sec:Ftheorylimit}}

In this final subsection we now turn to the discussion of the F-theory limit of sending the fiber volume~$v^0$ to zero.
Our aim is to show how the infinite charge orbit obtained in section~\ref{sec:infinite_elliptic} corresponds 
to the Kaluza-Klein tower associated to the circle reduction in the F-theory side. 
Note that the F-theory limit corresponds to decompactifying the circle~$r\rightarrow \infty$ 
while keeping~$\cV_b$ finite. In this limit we recover the 6D effective theory of F-theory compactified on a Calabi-Yau threefold
with all 6D fields not taken to any limit in stark contrast to the limits discussed in subsection~\ref{lv_Mtheory}.

To begin with we discuss the F-theory limit in more detail and the map to the M-theory side. This limit corresponds to sending~$r \rightarrow \infty$ while keeping all~$j^\alpha$  and~$\cV_{\rm b}$ fixed. 
For convenience, let us assume that the radius diverges as~$r\sim \lambda\rightarrow \infty$. From~\eqref{rj-va-map} and~\eqref{cV-in_tildeva} we find that it is implemented in the 5D M-theory 
moduli space spanned by the coordinates~$v^0, v^\alpha$ as 
\begin{equation}
\label{Flimit}
v^\a \sim \lambda^{2/3}\rightarrow \infty, \qquad v^0 \sim \lambda^{-4/3}\rightarrow 0\ .
\end{equation}
In other words, \textit{all}~$v^\a$ become large while~$v^0$ vanishes at a rate~$v^\a/v^0\sim \lambda^2\rightarrow \infty$. This also implies that the overall volume~$\cV$ in Planck units stays finite and so does the volume of the base~$\cV_b$ in string units on the F-theory side. 
From the definition of~$j^\a$ in~\eqref{j^a}, one then finds that~$v_\bb^\a$ scales as
\begin{align}
v_\bb^\a \sim  \sqrt{v_0} \, v^\a 
\end{align}
in the~$r\to \infty$ limit. This is perfectly consistent with~\eqref{volumes}.

Our next task is to compute the infinite charge orbit in the limit~\eqref{Flimit} of the M-theory geometry. 
Note that the limit~\eqref{Flimit} is just a special case of the limits studied in subsection~\ref{sec:smallfiber}. 
In fact, we can use the Fourier-Mukai transform introduced in~\eqref{S-explicit} and~\eqref{t->1/t} 
to transfer the orbits at~$v^0 \rightarrow \infty$ to~$v^0 \rightarrow 0$ by sending~$v^0 \rightarrow 1/v^0$. 
Furthermore, since we 
know the precise growth of~$v^\a$ and~$v^0$, we can infer which large volume limit we need to consider.  
To avoid confusion, let us call the large volume variable~$\tilde v^0 = 1/v^0$. Then~\eqref{Flimit} corresponds  
to the large volume limit 
\begin{equation}\label{tildev_grows_faster}
v^\a \sim \lambda^{2/3}\rightarrow \infty, \qquad \tilde v^0 \sim \lambda^{4/3}\rightarrow \infty\ .
\end{equation}
In other words, the fiber volume grows faster than all coordinates~$v^\alpha$. 
This determines the relevant charge orbit at large volume as discussed in subsection~\ref{sec:infinite_elliptic}.
Furthermore, we can employ the transformation~\eqref{QF=SQ}  to transfer the orbit 
to small fiber volume yielding 
\begin{equation}\label{QF-orbit}
\mathbf Q_{F} 
= \Big( 0,\ 0,\ \base^{\alpha\beta} q_\alpha^{(2)},0,\ -m_0 q_0^{(2)}-\textstyle \sum_{\alpha} m_\alpha q^{(2)}_{\alpha}+\frac12 (K^\alpha-\base^{\alpha\beta}K_{\beta\beta 0})q_\alpha^{(2)},\ q_0^{(2)}\Big)^{\rm T}\ ,
\end{equation}
which is a special case of the orbit given in~\eqref{QF-sec35}.
It was a central result of subsection~\ref{sec:infinite_elliptic} that one is allowed to set~$q_\alpha^{(2)} = 0$, for all~$\alpha=1,\ldots, h^{1,1}(B_2)$ and take~$q_0^{(2)} \neq 0$ to generate an infinite orbit becoming massless in the limit~\eqref{tildev_grows_faster} and valid for any Calabi-Yau. 
Making this choice in~\eqref{QF-orbit} one finds 
\begin{equation}\label{special_QF_2}
\mathbf Q_{F} 
= \Big( 0,\ 0, 0 ,0,\ -m_0 q_0^{(2)},\ q_0^{(2)}\Big)^{\rm T}\ .
\end{equation}
Before turning to the interpretation of this orbit, let us stress that it does not satisfy the conditions outlined in subsection~\ref{Kahler_charge_orbits} in the small fiber volume regime, since in certain cases there is no monodromy operator that can generate an infinite massless orbit in this regime. The orbit is rather transferred from the large volume regime and involves an~$Sl(2,\mathbb{Z})$ rotation of the charges (recall figure~\ref{fig_lv_sv} in which the F-theory limit corresponds indeed to small fiber and large base volume).

Finally, let us interpret the orbits~\eqref{QF-orbit} and~\eqref{special_QF_2}. To begin with, we note that, as in the previous subsection,
the orbits are actually Type IIA orbits and hence their entries correspond to charges of D$p$-branes. Connecting the M-theory 
setting of this section with the Type IIA orbit, we compactify on a further~$\hat S^1$. The 
last entry of the orbits corresponds to D0-brane charge in Type IIA and lifts to 
KK-momentum of an M2-brane state in M-theory. In fact, the orbits also admit non-trivial M2-brane charge as soon as 
$q^{(2)}_I \neq 0$ and thus describe 
M2-branes on the specified curves. 
The very special orbit~\eqref{special_QF_2} has in addition to D0-charge only D2-charge corresponding to a 
brane wrapped on the curve~$-m_0 q_0^{(2)} \cC^0$. In M-theory one thus finds an M2-brane tower wrapping multiple times 
the elliptic fiber and having a certain KK-momentum 
around~$\hat S^1$. Clearly, we can also proceed for more general orbits in~\eqref{QF-orbit} that admit D4-brane charge. This indicates that M5-branes wrapped on~$D_\a \base^{\alpha\beta} q_\b^{(2)}$ and~$\hat S^1$ will be relevant in the limit. 

In the next step one has to dualize the M-theory states to F-theory. Following the standard M/F-duality an M2-brane state on the elliptic fiber dualizes to a fundamental Type IIB string with KK-momentum along the circle~$S^1$ connecting the 
5D M-theory setting with the 6D F-theory setting. This implies that the orbit~\eqref{special_QF_2} labels the KK-tower of the 
6D fields. To see this explicitly we need to change into the basis of two-forms as discussed around~\eqref{KK2form}. The Kaluza-Klein vector associated to the~$S^1$ circle reduction comes from expanding~$C_3$ as~$C_3=A^{\rm KK}\wedge \check\omega_0$. 
The charge of the infinite orbit under the KK vector~$A^{\rm KK}$ is then given by
\begin{equation}
\int H_F\wedge \check\omega_0= \mathbf{Q}_{\rm F}^{I}\ \pairing_{IJ} \ \big (\delta^J_1+\mfrac 12 K^\alpha\delta^J_\alpha\big )= -m_0 q_0^{(2)}
\ ,
\end{equation}
where~$H_F$ is an even form with coefficients~$\mathbf{Q}_{\rm F}^{I}$.
Analogously, it is not hard to check that the charge under any of the other 5D gauge 
boson~$A^\a$ is zero since~$(Q_F \cdot \pairing)_J\delta^J_\alpha=0$.
Therefore, the tower of states only differ by their charge under the KK photon associated to the circle compactification of the 6D F-theory effective action to five dimensions. 
More generally, for the orbit~\eqref{QF-orbit} one has to also follow M5-branes through the M/F-duality. Since 
these M5-branes wrap the elliptic fiber they dualize to D3-branes wrapping a curve in~$B_2$. These D3-branes yield 
string states in the 6D effective theory which couple to the tensor fields. This matches with the fact that in 5D they are charged 
under~$A^\a$, i.e.~the vector arising from the 6D tensor fields~$\hat B_\a$. We leave a more detailed analysis of these strings for the future. At the moment, we conclude this section by remarking the identification of the Kaluza-Klein tower of the F-theory circle with the universal infinite massless charge orbit in the M-theory geometry.

\section{Conclusions} \label{conclusions}

In this paper we have investigated the Swampland Distance Conjecture, and the associated notion of 
emergence of infinite field distances, in the context of K\"ahler moduli spaces of 
Calabi-Yau manifolds. For the conjecture to hold there should exist an infinite tower of states near every 
infinite distance locus of the moduli space whose mass decreases exponentially fast in terms of the proper geodesic 
field distance to this locus. The proposal of \cite{Grimm:2018ohb} is to identify this tower with an infinite orbit of states charged under the discrete infinite symmetries which are part of the duality group of the string compactification. More concretely, this discrete symmetry corresponds to the monodromy transformation that the mirror period vector undergoes when circling the infinite distance locus. As these monodromies enhance to a continuous transformation at infinite distance, the infinite tower can then be understood as a quantum gravity obstruction to restore a global symmetry. We have 
also further elucidated the more speculative proposal of \cite{Grimm:2018ohb} that quantum corrections from integrating out the SDC tower are responsible for generating the infinite field distance itself.  

It was explained in reference \cite{Grimm:2018cpv} that powerful mathematical orbit theorems and the theory of limiting mixed Hodge structures allows one to classify the infinite distance loci and construct the massless infinite charge orbits in the complex structure moduli space of Calabi-Yau threefolds in complete generality. While this gives a general proof of the existence 
of an orbit under the stated assumptions, the constructions presented in \cite{Grimm:2018cpv} are technically involved and hard to apply to explicit examples. In this paper, we have shown that the same mathematical technology can be used to state the masslessness and infiniteness conditions as vector equations that then can be solved for concrete examples.
In particular, our approach allowed us to construct the infinite charge orbits at the infinite distance loci of K\"ahler moduli spaces. In the large volume regime, the generic form of the log-monodromies and symplectic form is fully determined by the topological data of Calabi-Yau manifold, namely its intersection numbers and Chern classes. We have argued that one can thus classify the possible singularity types and possible singularity enhancement chains corresponding to partial decompactification limits entirely 
using the intersection numbers.
With these at hand, we then identified the infinite charge orbits that are massless when approaching any infinite distance point
in the large volume regime. We provided the general form of the orbit, in terms of the singularity type, valid for any Calabi-Yau threefold and identified the corresponding D-brane states. This provides yet another strong piece of evidence for the SDC in the context of String Theory. 

Having discussed the general charge orbit in the large volume regime, we then further focused our study to 
the cases in which the Calabi-Yau manifold is elliptically fibered. The special intersection pattern of these geometries 
allowed us to give a detailed account of the arising large volume charge orbits. In particular, we were able to identify 
a universal orbit that is generically massless if the volume of the elliptic fiber is send to infinity.  
We then further exploited the geometry of elliptic fibrations, to ague that the orbits from the large volume regime can be transferred to regime of small fiber and large base volumes. This is done  by applying two T-dualities along the elliptic fiber and  a so-called operation Fourier-Mukai transformation on the D-brane charges. In this manner, we are able to obtain  infinite charge orbits becoming massless at the small-fiber regime. We stress that this is the first construction that goes beyond analyzing the SDC in a local region of the moduli space (see also \cite{Gonzalo:2018guu} for a very recent analysis of the SDC beyond perturbative level also using modular symmetries). It explicitly realizes the transfer of a charge orbit from a region in moduli space which allows for a local construction to a different regions of the moduli space where no such local construction 
is possible.

It is important to stress that, as our above constructions show, the infinite charge orbit does not always have the interpretation of a Kaluza-Klein tower, even if this is the naive candidate for an infinite tower becoming massless at large volume. In fact, depending on the particular string theory setup, it can also correspond to particles or strings coming from wrapping branes. If we consider Type IIA compactified in a Calabi-Yau threefold, the charge orbit at large volume consists of particles arising from bound states of D0-D2 branes wrapping certain two-cycles, which lift to M2-brane states in M-theory. Even if they get heavy at the large volume limit, they are exponentially light compared to the Planck scale and hence become massless if we force the Planck mass to remain finite. There are, therefore, two equivalent ways to avoid the restoration of the global symmetry, either gravity decouples ($\mp \rightarrow \infty$) or the infinite tower of states becomes massless leading to an exponential drop-off of the quantum gravity cut-off. For the case of Type IIA, this global symmetry corresponds to an axionic continuous shift symmetry that is lifted to a one-form global symmetry in M-theory. 

In the second part of this paper we also analysed the F-theory interpretation of the infinite massless charge orbit at the different infinite distance loci. For the large volume limits each charge orbit corresponds to a 6D string wrapping the F-theory circle to five dimensions. Each such 6D string in turn arises from a D3-brane in Type IIB, which is wrapping a non-trivial two-cycles in the base of the elliptic fibration whose volume is not sent to infinity. The identification of this string with an infinite orbit in M-theory makes manifest the fact that the string should count as infinitely many different particles. This suggests a potential application of these infinite charge orbits beyond the SDC, as a promising tool to count the number of different massless excitations of extended objects in F-theory. 
We then investigate the interpretation of the infinite massless charge orbits at the small fiber regime, which maps to decompactifying the additional circle of the F-theory compactification. We find that the infinite massless charge orbits at the F-theory limit always differ by their charge under the KK photon of the F-theory circle, hinting the existence of the extra dimension. In particular, we show that there always exists a universal infinite orbit regardless of the specific intersection numbers of the Calabi-Yau, that maps to the Kaluza-Klein tower of the 6D fields in F-theory. This provides a geometric realization of the KK tower in terms of an infinite massless charge orbit in M-theory.  We also get that there could be other infinite towers identified with 6D strings coming from M5-branes, whose analysis is left for future work.

Last but not least, we pay special attention to whether the infinite field distance can emerge from integrating out the infinite tower of states. First, we present a general field theory computation to show that, as long as the tower gets compressed as we move in the moduli space, quantum corrections from integrating out the tower up to its species bound will generate the infinite field distance. Remarkably, they will generate a logarithmic divergence of the field distance as a function of the mass of the tower, regardless of the specific form of the mass, and yielding the exponential mass behavior required by the SDC. We find that the condition for these quantum corrections to dominate over the classical piece in the IR matches with the constraint on the mass spectrum imposed by the Scalar Weak Gravity Conjecture \cite{Palti:2017elp}. If we apply this reasoning to a Kaluza-Klein circle reduction in field theory, the species bound associated to the KK tower turns out to be the Planck mass of the higher dimensional theory. However, quantum corrections from the KK tower can only account at most for part of the infinite field distance as the radius goes to infinity. The situation changes when considering similar setups in string theory. As mentioned, the infinite tower of states becoming massless at large volume of Type IIA Calabi-Yau compactifications consists of D0-D2 branes which could in fact completely generate the infinite field distance. Notice that this means that the field metric in the K\"ahler moduli space, and consequently the intersection numbers and topological discrete data of the Calabi-Yau, would be emergent from integrating out these D0-D2 bound states. Finally, the emergence of the classical quantities in the M-theory reduction from integrating out states has also a clear interpretation in the context of the M/F-theory duality. There, it is known \cite{Cvetic2013,Bonetti:2013ela} that some of the Chern-Simon terms arising in the M-theory dimensional reduction at classical level can only be recovered in the F-theory side upon taking into account quantum corrections from integrating out the KK tower associated to the F-theory circle. These Chern-Simon terms are related to the field metric by supersymmetry, so at least part of the metric yielding the infinite field distance in the F-theory limit arises form integrating out the KK tower. While this nicely supports the idea of emergence in this context, it is only a first step to show that the infinite distance entirely emerges from integrating out these infinite towers. To confirm the emergence 
conjecture one likely needs to keep track of any possible tower of states becoming massless in this limit as they might all contribute to generate the full divergence of the distance.  

There are also a few further points that are interesting to address in future work. First, we have assumed that the K\"ahler cone is simplicial, so the natural next step is to remove this assumption and generalize the classification of singularities and charge orbits to non-simplicial cones. Secondly, while we have focused on identifying explicit universal charge orbits that are present for any Calabi-Yau manifold at the different types of infinite distance singularities, the structure of all possible existing massless charge orbits is more complicated and can depend on the topological discrete data of the manifold. It would be interesting to perform a detailed study of all existing orbits and their microscopic interpretation in string theory, as well as their possible role in the emergence of the infinite distance. Lastly, we have not shown yet if the charge orbits are populated by physical states as we approach the singular point. The monodromy transformation guarantees the presence of an infinite number of physical states at the singularity as long as a single charge of the orbit is populated. However, the question remains how the stability of the states changes 
when approaching the singularity. It would be then important to realize an analysis of possible walls of marginal stability, as performed in \cite{Grimm:2018ohb}, to check that the number of physical states populating the tower indeed increases exponentially as we approach the singularity, as the species bound and the idea of emergence suggest. 

Finally, in this paper we have focused on the Swampland Distance Conjecture, but recent works are pointing to an interesting emerging network of relations between the different Swampland Conjectures (see \cite{Ooguri:2018wrx} for a relation with the de Sitter swampland conjecture \cite{Obied:2018sgi}). In particular, the above infinite distance limits can also correspond to weak coupling limits for the gauge bosons completing the~$N=2$ vector multiplets. In that case, the infinite charge orbit would also correspond to the states satisfying the Weak Gravity Conjecture \cite{ArkaniHamed:2006dz}, as discussed in \cite{Grimm:2018ohb,Lee2018,Lee2019}. We leave for future work a more detailed analysis of their charge to mass ratio, which can help to properly define the WGC in the presence of both scalar and gauge fields.

\vspace{1cm}

\noindent
{\bf \large Acknowledgments}

\noindent
It is a pleasure to thank Michael Fuchs, Chongchuo Li, Miguel Montero, Eran Palti, and Thorsten Schimannek for valuable
discussions and correspondence. PC would like to thank the ITP KULeuven and the IPhT Saclay for their hospitality during completion of the final part of this work. IV is supported by the Simons Foundation Origins of the Universe program (Modern Inflationary Cosmology collaboration).

\appendix

\section{Constructing the massless infinite charge orbits}
\label{app:orbits}

In this appendix, we derive the masslessness conditions \eqref{masslessness} and \eqref{masslessnessII} presented in the main text, and explicitly construct orbits satisfying them, as well as the infiniteness condition~\eqref{cond_orbit}.

As explained in the main text, since 
\begin{equation}
m(\mathbf Q) = \vabs{Z(\mathbf Q)} \leq \norm{\mathbf Q} \sim \norm{\qo},
\end{equation} 
having $\norm{\qo} \to 0$ is \emph{sufficient} to ensure masslessness of the BPS states with charge vector $\mathbf Q$. Note that since $m(\qo) = \abs{Z(\qo)} \le \norm{\qo}$, the states corresponding to $\qo$ is also massless.
In \cite{Grimm:2018cpv} it was established that for a singularity $t^i \to \infty, \ i=1,\ldots,\infty$ a $\qo \in W_{l_1} (\N{1}^-) \cap W_{l_2} (\N{2}^-) \cap \ldots \cap W_{l_n} (\N{n}^-)$, where the $l_i$'s are the smallest values for which this is true, has a vanishing norm if the following condition is satisfied
\begin{equation}
l_n <3 \qqtext{and} l_1, \ldots, l_{n-1} \le 3 \label{van_norm}
\end{equation}
The conditions for a vector to belong to certain $W_l = \bigoplus_{p+q\le l} I^{p,q} $ depend on the $I^{p,q}$ of the considered singularity. The $I^{p,q}$ naturally split into primitive parts $P^{p,q}$ and non-primitive parts, of the form $N^k P^{p,q}$. This decomposition is given explicitly for the different singularity types in Table \ref{Ipq}, from which one can also read the conditions for $\qo$ to belong to $W_2$ or $W_3$. We refer the reader to \cite{Grimm:2018cpv,Grimm:2018ohb} for more details.
\begin{table}[h!]
	\tabcolsep=4mm
	\arraycolsep=1mm
	\begin{center}
		\begin{tabular}{@{}l@{\hspace*{-1mm}}cll@{}}	\toprule 
			Sing. type & $I^{p,q}$ decomposition & $\qo \in W_3$ & $\qo \in W_2$ \\	\midrule 
			I & $\begin{array}{ccccccc}
			&   &         &     0     &         &   &         \\
			&   &    0    &           &    0    &   &         \\
			& 0 &         &  P^{2,2}  &         & 0 &         \\
			P^{3,0} &   & P^{2,1} &           & P^{1,2} &   & P^{0,3} \\
			& 0 &         & N P^{2,2} &         & 0 &         \\
			&   &    0    &           &    0    &   &         \\
			&   &         &     0     &         &   &
			\end{array}$ & $ \qo = \v $ & $ \qo = N \u $ \\ \midrule
			II & $\begin{array}{ccccccc}
			&           &         &     0     &         &           &   \\
			&           &    0    &           &    0    &           &   \\
			&  P^{3,1}  &         &  P^{2,2}  &         &  P^{1,3}  &   \\
			0 &           & P^{2,1} &           & P^{1,2} &           & 0 \\
			& N P^{3,1} &         & N P^{2,2} &         & N P^{1,3} &   \\
			&           &    0    &           &    0    &           &   \\
			&           &         &     0     &         &           &
			\end{array} $ & $\qo = \v $ & $ \qo = N \u $\\ \midrule 
			III & $ \begin{array}{ccccccc}
			&   &                          &     0     &                          &   &   \\
			&   &         P^{3,2}          &           &         P^{2,3}          &   &   \\
			& 0 &                          &  P^{2,2}  &                          & 0 &   \\
			0 &   & P^{2,1} \oplus N P^{3,2} &           & P^{1,2} \oplus N P^{2,3} &   & 0 \\
			& 0 &                          & N P^{2,2} &                          & 0 &   \\
			&   &       N^2 P^{3,2}        &           &       N^2 P^{2,3}        &   &   \\
			&   &                          &     0     &                          &   &
			\end{array} $ & $ \qo = \v + N \u $ & 
			$ \qo = N \w$ 
			\\ \midrule
			IV & $ \begin{array}{ccccccc}
			&   &         &           P^{3,3}            &         &   &   \\
			&   &    0    &                              &    0    &   &   \\
			& 0 &         &   P^{2,2} \oplus N P^{3,3}   &         & 0 &   \\
			0 &   & P^{2,1} &                              & P^{1,2} &   & 0 \\
			& 0 &         & N P^{2,2} \oplus N^2 P^{3,3} &         & 0 &   \\
			&   &    0    &                              &    0    &   &   \\
			&   &         &         N^3 P^{3,3}          &         &   &
			\end{array} $ & 
			$ \qo = \v + N \x  $ 
			& 
			$ \qo =  N \w + N^2 \u $ 
			\\
			\bottomrule
		\end{tabular}
		\caption{We present for each singularity type the explicit splittings of the $I^{p,q}$ in term of the primitive subspaces $P^{p,q}$, namely $I^{p,q} = \oplus_{i\ge 0} \, N^i P^{p+i, q+i}$. From these one can read off the conditions for $\qo \in W_2$ or $\qo \in W_3$, which are then given in the third and fourth column, where the vector $\u$ is unconstrained, while the vectors $\v$, $\w$ and $\x$ satisfy $N \v = 0$, $N^2 \w =0 $ and $N^3 \x =0$.}
		\label{Ipq}
	\end{center}
\end{table}

\paragraph{Masslessness conditions}
Applying these conditions to \eqref{van_norm}, we find that the conditions for the  seed vector $\qo$ to be massless are those stated in the main text, namely \eqref{masslessness} and \eqref{masslessnessII}, which we recall here again for convenience
\begin{equation}\label{massl_cond}
\tabcolsep=4mm
\begin{tabular}{@{}ll@{\hspace{10mm}}ll@{}}
	$\type{i}$ & $\qo$                & $\type{n}$ & $\qo$                      \\ \midrule [0.2pt]
	II         & $ \v_i $             & II         & $\N{n} \u_n $              \\
	III        & $ \v_i + \N i \u_i$  & III        & $\N n \w_n $               \\
	IV         & $ \v_i + \N i \x_i $ & IV         & $\N n \w_n + \N n^2 \u_n $
\end{tabular}
\end{equation}
where $\N i \v_i =0$, $\N i^3 \x_i =0$ and $\N n^2 \w_n = 0 $. 

\paragraph{Infiniteness conditions} 
In addition, we recall the condition \eqref{cond_orbit} for the orbit to be generated 
\begin{equation}\label{inf}
\N{J^*} \qo \neq 0 \qquad \text{for some $J^*= 1,\ldots,\hoo{Y_3}$}
\end{equation}

We now proceed to satisfy those conditions, that is, to explicitly give the vectors 
$\v_i$, $\u_i$, $\x_i$, $\u_n$ and $\w_n$
such that eqs.~\eqref{massl_cond}-\eqref{inf} hold.
Before specializing to the different singularity types, let us recall here the explicit form of the matrix $\N i$ and its powers (given in eqs. \eqref{Nn} and \eqref{Nn2} for $i=n$)
\arraycolsep=2pt
\begin{equation}
\N{i} = \left(
\begin{array}{cccc}
0                   & 0                  & 0            & 0\\
-\sum_{a}^i\delta_{aI}        & 0                  & 0            & 0\\
-\frac{1}{2} \KKbis{I}{i} & - \KK{IJ}{i}           & 0            & 0\\
\frac{1}{6} \KKbis{}{i}  & \frac{1}{2} \KK{JJ}{i} & -\sum_{a}^i\delta_{aJ} & 0
\end{array}\right)\, \quad 
\N{i}^2  
= \left(
\begin{array}{cccc}
0    &    0    & 0 & 0 \\
0    &    0    & 0 & 0 \\
\KK{I}{i} &    0    & 0 & 0 \\
0    & \KK{J}{i} & 0 & 0
\end{array}\right) \quad
\N{i}^3
= \left(
\begin{array}{cccc}
0     & 0 & 0 & 0 \\
0     & 0 & 0 & 0 \\
0     & 0 & 0 & 0 \\
- \KK{}{i} & 0 & 0 & 0
\end{array}\right).
\end{equation}
where we defined $\KKbis{I}{i} = \sum_{a=1}^i \K{aaI}$ and $\KKbis{}{i} = \sum_{a=1}^i \K{aaa}$. With these at hand, we find their action on a generic vector $\q = (\q^6,\q^4_I,\q^2_I,\q^0)^\T$ --- a convention will also adopt for the vectors $\u$, $\v$, $\w$ and $\x$ throughout this appendix --- to be
\begin{equation} \label{Nq}
\N{i} \q = 
\begin{pmatrix}
0\\
-\sum_a^i\delta_{aI} \, \q^6\\
-\frac{1}{2} \KKbis{I}{i} \, \q^6 - \KK{IJ}{i} \, \q^{4,J} \\
\frac{1}{6} \KKbis{}{i} \q^6 + \frac{1}{2} \KK{JJ}{i} \, \q^{4,J} -\sum_a^i \, \q^{2,a}
\end{pmatrix}, \quad 
\N{i}^2 \q =\begin{pmatrix}
0 \\0 \\  \KK{I}{i} \, \q^6 \\ \KK{I}{i} \, \q^{4,I}
\end{pmatrix}, \quad 
\N{i}^3 \q =\begin{pmatrix}
0 \\0 \\ 0 \\  - \KK{}{i} \, \q^6 
\end{pmatrix}\,.
\end{equation}
This will allow us to translate the conditions in \eqref{massl_cond} into conditions on the components of the vectors.
The analysis depends on the type of the last singularity in the considered chain, i.e. $\type{n}$ in \eqref{t>inf_chain}.
We now specialize to the different possible singularity types.

\subsection{$\type{n} = \rII$}

The first and simplest situation is when $\type{n} = \rII$, where the masslessness conditions are, as can be read from  \eqref{massl_cond},
\begin{subequations}
\begin{align}
& \qo^\rII = \v_i \qtext{where} \N i \v_i = 0 \qquad  \text{for} \quad i \le n \,, \label{II2}\\
&\qo^\rII = \N{n} \u_n \,. \label{II1}
\end{align}
\end{subequations}
Eq. \eqref{II1} implies
\begin{equation}\label{qoII}
\qo^\rII =   \begin{pmatrix}
0\\
-\sum_a^n\delta_{aI} \, \u_n^6\\
-\frac{1}{2} \KKbis{I}{n} \, \u_n^6 - \KK{IJ}{n} \, \u_n^{4,J} \\
\frac{1}{6}\KKbis{}{n} \, \u_n^6 + \frac{1}{2} \KK{JJ}{n} \, \u_n^{4,J} -\sum_a^n \, \u_n^{2,a}
\end{pmatrix}.
\end{equation}
Acting on this $\qo^\rII$ with $\N i$ we find 
\begin{equation}
\N{i} \qo^\rII = \begin{pmatrix} 
0 \\ 0 \\  \sum_{a}^i \KK{aI}{n} \u_n^6 \\
- \mfrac 12 \sum_{a}^i  \KK{aa}{n} \u_n^6 - \sum_{a}^i (-\frac{1}{2} \KKbis{a}{n} \, \u_n^6 - \KK{aI}{n} \, \u_n^{4,I})
\end{pmatrix}.
\end{equation}
Since $\type{n} = \rII$, one has $\KK{}{n} = \KK{I}{n} =0$ which implies
\begin{equation}
\KK{aa}{n} = \KKbis{a}{n} = \KK{aI}{n} = 0 \Qtext{for} a,b \le n \,,
\end{equation}
such that the condition \eqref{II2} is automatically satisfied. This means that $\qo^\rII$ in \eqref{qoII} is the generic form of a massless seed vector. On the other hand The infiniteness condition \eqref{inf} gives 
\begin{equation}\label{infII}
\N{J^*} \qo^\rII = \begin{pmatrix}
0 \\ 0 \\  \sum_{a}^{J^*} \KK{aI}{n} \u_n^6 \\
- \mfrac 12 \sum_{a}^{J^*}  \KK{aa}{n} \u_n^6 - \sum_{a}^{J^*} (-\frac{1}{2} \KKbis{a}{n} \, \u_n^6 - \KK{aI}{n} \, \u_n^{4,I})
\end{pmatrix} \ne 0 \,.
\end{equation}
Since $\rk \KK{IJ}{n} \ne 0$, there are some $I^*$ and $J^*$ such that $\KK{I^* J^*}{n} >0$, eq. \eqref{infII} can be satisfied, both if $\u_n^6 \neq 0$ or $\sum_a^{J^*} \KK{aI}n \u_n^{4,I} \neq 0$, in particular one can have a solution with $\u_n^6 =0$.\footnote{For instance choosing $\u_n^{4,I} = 1$ for all $I$ a possible solution, but it is of course not the only one.} 
As mentioned in the main text, the last entry of $\qo$ plays no role and can safely be set to zero, here by choosing $\sum_a^n \u_n^2 = \frac 12 \KK{JJ}{n} \u_n^{4,J}$. Making those choices and renaming $\o^I = - \u_n^{4,I}$, we find 
\begin{equation}
\qo^\rII = \Big(  0,0, \KK{IJ}{n} \o^J,0\Big)^\T\,.
\end{equation}

\subsection{$\type{n} = \rIII$}

The next situation is $\type{n} = \rIII$, where the masslessness conditions are, as can be read from \eqref{massl_cond},
\begin{subequations}\label{III}
\begin{alignat}{2}
& \qo^\rIII = \v _i   \qtext{where} \N i \v_i = 0 &&  i < n_\rIII \,, \label{III2}\\
& \qo^\rIII = \v _j + \N j \u_j  &&   n_\rIII \le j < n \label{III3} \,, \\
& \qo^\rIII = \N n \w_n \qtext{where} \N n ^2 \w_n =0 \,, \quad && \label{III1} 
\end{alignat}
\end{subequations}
where $n_\rIII$ is the first place where a type $\rIII$ singularity occurs.
Equations \eqref{III} lead to 
\begin{align}
\label{condIII}
\qo^\rIII &=  \begin{pmatrix}0\\ \v_i^{4}{}_I\\ \v_i^{2}{}_I \\\v_i^0
\end{pmatrix}
= \begin{pmatrix} 0\\
\v_j^4{}_I-\sum_a^j\delta_{aI} \, \u_j^6\\
\v_j^2{}_I-\frac{1}{2} \KKbis{I}{j} \, \u_j^6 - \KK{IJ}{j} \, \u_j^{4,J} \\
\v_j^0  +\frac{1}{6} \KKbis{}{j} \, \u_j^6 + \frac{1}{2} \KK{JJ}{j} \, \u_j^{4,J} -\sum_a^j \, \u_j^{2}{}_a
\end{pmatrix} =  \begin{pmatrix}0\\0\\
- \KK{IJ}{n} \, \w_n^{4,J} \\
\frac{1}{2} \KK{JJ}{n} \, \w_n^{4,J} -\sum_a^n \, \w_n^{2,a}
\end{pmatrix}   \,,
\end{align}
where $i < n_\rIII$ and $n_\rIII \le j < n$, and the components of $\w_n $ and $\v_i$ satisfy, for all $i<n$,
\begin{align}
\KK{I}{n} \w_n^{4,I} &= 0 \label{w4} \\
\KK{IJ}{i} \v_i^{4,J} & = 0 \label{v4=0} \\
\KK{II}{i} \v_i^{4,I} & = 2 \sum_{a=1}^{i}  \v_i^{2}{}_a \label{v4=v2}
\end{align}
From \eqref{condIII} we must impose $\qo^{\rIII,4} =0 $, such that 
\begin{subequations} \label{v4}
	\begin{alignat}{2}
\v_i^{4,I} & =0   && \Qtext{for} i < n_\rIII  \\
\v_i^{4,I} & = \sum_a^i \delta_{aI} \u_i^6   && \Qtext{for}  n_\rIII \le i <n
\end{alignat}
\end{subequations}
Condition \eqref{v4=0} then implies
\begin{equation}
\u_i^6 \,  \KK{I}{i}   = 0   , \qquad n_\rIII \le i < n \,,
\end{equation}
which leads to $\u_i^6 =0$ for $n_\rIII \le i < n$, since for a type III singularity $\KK{I}{n} \ne 0$. Eq. \eqref{v4} then implies that $\v_i^{4,I} =0 $ for all $ i $'s.
Condition \eqref{v4=v2} then becomes for $ n_\rIII \le i < n $
\begin{equation}
\sum_a^i \KK{aI}{n} \w_n^{4,I}  = \KK{I}{i} \, \u_i^{4,I} \,,
\end{equation}
which can always be satisfied since $\u_i^{4}$ is arbitrary and $\KK{I}{n}$ is non-vanishing. So it does not constrain $\w_n^4$. As before we choose $q^{(0)}$ to vanish by an appropriate choice of $\v_i^0$, $\u_i^2$ and $\w_n^2$ and rename $\o^I = - \w_n^{4,I}$ such that 
\begin{equation}
\qo^\rIII = \Big(  0,0, \KK{IJ}{n} \o^J,0\Big)^\T\,,
\end{equation}
together with the conditions \eqref{v4=v2} for $ i < n_\rIII $ and \eqref{w4} that now read 
\begin{alignat}{2}
\KK{iI}{n} \, \o^I &  = 0  \qquad  &&   i  < n_\rIII  \,, \label{condIIIfin1}\\
\sum_{a=1}^{n} \KK{aI}{n} \, \o^I & = 0 \,, && \label{condIIIfin2}
\intertext{while the condition \eqref{inf} for the orbit to be generated}
\sum_{a=1}^{J^*} \KK{aI}{n} \, \o^I & \ne 0 && \text{for some} \quad  J^* \,. \label{infIII}
\end{alignat}
An easy way to satisfy these equations is to choose 
\begin{equation} \label{omegaI}
\o^I = \begin{cases}
1 \quad & I \le n\\
0 \quad & I > n
\end{cases} \,,
\end{equation}
which leads to $q^\two_I = \KK{I}{n}$; this is non-vanishing for a type III singularity, meaning that indeed \eqref{infIII} is satisfied, and since in addition $\KK{}{n} = 0$ for a type III, one has $\KK{iI}{n} =0$ for all $i \le n$, such that \eqref{condIIIfin1} and \eqref{condIIIfin2} are also satisfied.

\subsection{$\type{n} = \rIV$}

Finally, when $\type{n} = \rIV$, the masslessness conditions are, as can be read from \eqref{massl_cond},
\begin{subequations}\label{IV}
	\begin{alignat}{2}
& \qo^\rIV = \v _i   \qtext{where} \N i \v_i = 0 && i < n_\rIII \label{IV2}\\
& \qo^\rIV = \v _i + \N i \u_i  &&   n_\rIII \le i < n_\rIV  \label{IV3}\\
& \qo^\rIV = \v _i + \N i \x_i \qtext{where} \N i^3 \x_i = 0 &&   n_\rIV \le i < n \label{IV4} \\
& \qo^\rIV = \N n \w_n + \N n^2 \u_n \qtext{where} \N n ^2 \w_n =0 \,, && \label{IV1}
\end{alignat}
\end{subequations}where $n_\rIV$ is the first place where a type $\rIV$ singularity occurs.
Equations \eqref{IV} lead to 
\begin{align}
\begin{split}
\label{condIV}
\qo^\rIV &=  \begin{pmatrix}0\\ \v_i^{4}{}_I\\ \v_i^{2}{}_I \\\v_i^0
\end{pmatrix}
= \begin{pmatrix} 0\\
\v_j^4{}_I-\sum_a^j\delta_{aI} \, \u_j^6\\
\v_j^2{}_I-\frac{1}{2} \KKbis{I}{j} \, \u_j^6 - \KK{IJ}{j} \, \u_j^{4,J} \\
\v_j^0  +\frac{1}{6} \KKbis{}{j} \, \u_j^6 + \frac{1}{2} \KK{JJ}{j} \, \u_j^{4,J} -\sum_a^j \, \u_j^{2}{}_a
\end{pmatrix} \\
& =
\begin{pmatrix} 0\\ \v_k^{4}{}_I \\
\v_k^{2}{}_I - \KK{IJ}{k} \, \x_k^{4,J} \\
\v_k^0 + \frac{1}{6} \KKbis{}{k} \, \x_k^6 + \frac{1}{2} \KK{JJ}{k} \, \x_k^{4,J} -\sum_a^k \, \u_k^{2}{}_a
\end{pmatrix} 
=  \begin{pmatrix}0\\0\\
- \KK{IJ}{n} \, \w_n^{4,J} +\KK{I}{n} \u_n^6  \\
\frac{1}{2} \KK{JJ}{n} \, \w_n^{4,J} -\sum_a^n \, \w_n^{2,a}+ \KK{I}{n}  \u_n^{4,I} 
\end{pmatrix}   \,,
\end{split}
\end{align}
where $i < n_\rIII$, $n_\rIII \le j < n_\rIV$, and $n_\rIV \le k< n$  and as in the previous case the components of $\w_n $ and $\v_i$ satisfy, for all $i<n$,
\begin{align}
\KK{I}{n} \w_n^{4,J} &= 0 \,, \label{w4-IV} \\
\KK{IJ}{i} \v_i^{4,J} & = 0 \,,\label{v4=0-IV} \\
\KK{II}{i} \v_i^{4,I} & = 2 \sum_{a=1}^{i}  \v_i^{2}{}_a \,. \label{v4=v2-IV}
\end{align}
From \eqref{condIV} we must impose $\qo^{\rIII,4} =0 $ such that 
\begin{subequations} \label{v4-IV}
	\begin{alignat}{2}
	\v_i^{4,I} & =0   && \Qtext{for} i < n_\rIII  \qtext{and} n_\rIV \le i < n \,,  \\
	\v_i^{4,I} & = \sum_a^i \delta_{aI} \u_i^6   && \Qtext{for}  n_\rIII \le i <n_\rIV \,.
	\end{alignat}
\end{subequations}
Condition \eqref{v4=0-IV} then implies
\begin{equation}
\u_i^6 \,  \KK{I}{i}   = 0   , \qquad n_\rIII \le i < n \,,
\end{equation}
which leads to $\u_i^6 =0$ for $n_\rIII \le i < n_\rIV$, since for a type III singularity $\KK{I}{n} \ne 0$. Eq. \eqref{v4-IV} then implies that $\v_i^{4,I} =0 $ for all $ i $'s.
Relabeling $\x_i^4 = \u_i^4$ when $ n_\rIV \le n <n$, condition \eqref{v4=v2} then becomes for $ n_\rIII \le i < n $
\begin{equation}
\sum_a^i \KK{aI}{n} \w_n^{4,I} - \KK{a}{n} \u_n^6   = \KK{I}{i} \, \u_i^{4,I} \,,
\end{equation}
which can always be satisfied since $\u_i^{4}$ is arbitrary and $\KK{I}{n}$ is non-vanishing. So it does not constrain $\w_n^4$ or $\u_n^6$. As before we choose $q^{(0)}$ to vanish by an appropriate choice of $\v_i^0$, $\u_i^2$ and $\w_n^2$. And defining 
\begin{equation}
\o^I = \w^{4,I}_n
- \begin{cases}
\u_n^6 \qquad   & i \le n \\
0 \qquad & i > n
\end{cases} \ ,,
\end{equation}
we find 
\begin{equation} 
\qo^\rIV = \Big(  0,0, \KK{IJ}{n} \o^J,0\Big)^\T\,, \label{q0IV}
\end{equation} 
together with the conditions \eqref{v4=v2-IV} for $ i < n_\rIII $ and \eqref{w4-IV} that now read 
\begin{alignat}{2}
\KK{iI}{n} \, \o^I &  = 0  \qquad  &&  \qquad i  < n_\rIII \,, \label{condIVfin1}\\
\sum_{a=1}^{n} \KK{aI}{n} \, \o^I & = - \u_n^6 \KK{}{n} \,. && \label{condIVfin2}
\end{alignat}
Since $\KK{}{n}$ is non vanishing for a type IV singularity and $\u_n^6$ is arbitrary, \eqref{condIVfin2} can always be satisfied by an appropriate choice of $\u_n^6$ and does not put any further constrain on $\o^I$. So the only non-trivial masslessness constraint is \eqref{condIVfin1}, to be satisfied together with the condition \eqref{inf} for the orbit to be generated, that is, one needs to find a solution to
\begin{subequations} \label{system}
	\begin{alignat}{2}
\KK{iI}{n} \, \o^I &  = 0    &&  i  < n_\rIII \,, \label{cond1} \\
\sum_{a=1}^{J^*} \KK{aI}{n} \, \o^I  & \ne 0 \qquad && \text{for some} \quad  J^* \,. \label{cond2}
\end{alignat}
\end{subequations}
Of course if ${\sf Type\ A}_1 = \rIII$, i.e. $n_\rIII =1$, there is no condition~\eqref{cond1} and the state corresponding to the seed vector \eqref{q0IV} is automatically massless.
We thus need to show that it is possible to solve the system \eqref{system} when ${\sf Type\ A}_1 = \rII$. We will show this explicitly in the case where have only two moduli, and in the case of an elliptic fibration. We leave the general case for a future analysis, but point out that, the more moduli we have, the bigger becomes the orthogonal space to $\textstyle \sum_{i=1}^j \c K_{iI}^{(n)}$, such that it increases the room for solving the system~\eqref{system}.

\begin{itemize}
	\item \textbf{Two moduli}
	
	We first consider a case with two moduli, $v^1$ and $v^2$, and the associated enhancement chain
	\begin{equation}
	{\sf Type\ A}_1 + {\sf Type\ A}_2 \longrightarrow \type{2}.
	\end{equation}
	As mentioned above, we need $ {\sf Type\ A}_1 = \rII $ and, of course, $\type{2} = \rIV$, that is we have, from Table~\ref{Type_Table2},
	$ \KK{}{1} = \K{111} =0$ and $\KK{}{2} = \K{222} + 3\,\K{122} >0 $. The system~\eqref{system} then becomes
	\begin{align}\label{2moduli}
	\begin{split}
	q_1^\two = \KK{1J}{2} \o^J & = \K{122} \, \o^2 =0 \,, \\
	q_2^\two = \KK{2J}{2} \o^J & = \K{122} \, \o^1 + (\K{122}+ \K{222})\, \o^2 \ne 0  \,.
	\end{split}
	\end{align}
	It is always possible to find a solution to this system of equations. Indeed, there are 2 possibilities
	\begin{itemize}
		\item $\K{122} = 0$, in which case $\K{222} \ne 0$ and the system is satisfied with $\o^2 \ne 0$,
		\item $\K{122} \ne 0$, in which case the system is satisfied with $\o^2 = 0$ and $\o^1  \ne 0$. 
	\end{itemize}

	\item \textbf{Elliptic fibrations}
	
	We now turn to the case of an elliptic fibration, which is the most relevant for our analysis, in particular for sections~\ref{sec:infinite_elliptic}~--~\ref{sec:smallfiber} and~\ref{lv_Mtheory}~--~\ref{sec:Ftheorylimit}. We refer to section~\ref{sec:infinite_elliptic} for the notations and the possible enhancement chains. 
	Recall that we have the moduli $v^0$, $v^\a$'s, with corresponding singularities types ${\sf Type\ A}_0 =\rIV$ and ${\sf Type\ A}_\a \neq \rIV$.
	We show that we can always choose 
	\begin{equation}\label{q_a_0}
	q^\two_0 \neq 0 \qtext{and} q^\two_\a =0,
	\end{equation}
	which is actually stronger than eqs.~\eqref{system}.
	Using the intersection numbers~\eqref{int_numb_Kc}, we find for the charges in~\eqref{q0IV}
	\begin{subequations}
		\begin{align}
		q_\a^\two = \KK{\a J}{n} w^J & = \base_{\a\b} w^\b - (K_\a-\base_{\a}) w^0, \\
		q_0^\two  = \KK{0 J}{n} w^J& = (K_\a - \base_\a) (K^\a w^0 - w^\a ),
		\end{align}
	\end{subequations}
	where we defined $\base_{\a} =  \textstyle \sum_{\b} \base_{\a\b}$ and $K_{\a} =  \base_{\a\b}K^\b$. 
	Since $\base_{\a\b}$ can be inverted (and we denote the inverse by $\base^{\a\b}$), we can choose $w^\a = \base^{\a\b}(K_\b-\base_{\b}) \, w^0$, which yields 
	\begin{subequations}
		\begin{align}
		q_0^\two &= \textstyle \sum_{\a} ( K_\a -\base_{\a}) \, w^0, \\
		q_\a^\two & = 0.
		\end{align}
	\end{subequations}
	since $K_\a \le 0$ and $\base_{\a} >0$, we have $q_0^\two \ne 0 $ and equation~\eqref{q_a_0} holds, and therefore \eqref{system} as well.
	
	Finally, let us remark that we could also choose $w^\a =  K^\a \, w^0$, leading to
	\begin{subequations}
		\begin{align}
		q_0^\two  & = 0,\\
		q_\a^\two &= \base_{\a} \, w^0.
		\end{align}
	\end{subequations}
	However, this choice would only be compatible with \eqref{system}if there are no type II in the chain, i.e. if the first singularity is associated to either a coordinate $v^\a$ with $\eta_{\a\a} \ne 0$, or to~$v^0$.
\end{itemize}

\section{Fourier-Mukai transformation} \label{FM_appendix}

Let~$D^{\rm b}_\a$ be the divisors generating the K\"ahler cone of the base~$B_2$ while the dual basis of curves generating the Mori Cone is denoted by~$C'^\a$. For a Calabi-Yau threefold~$Y_3$ corresponding to an elliptic fibration over this base, we can define the curves
\begin{equation}
C^\a=E\, \pi^{-1}C'^\a, \quad \a=1,\dots,\hoo{B_2}
\end{equation}
where~$E$ is the zero-section of the elliptic fiber. A basis of the Mori cone of the Calabi-Yau is then given by~$\{C^I\}=\{C^0,C^\a\}$ where~$[C^0]$ is the class of the generic fiber. The K\"ahler cone is generated by the dual basis~$\{D_I\}=\{D_0,D_\a\}$ where
\begin{equation}
D_\a=\pi^*D^{\rm b}_\a \quad D_0=E+\pi^*c_1(B_2)
\end{equation}
such that~$D_I \cdot C^J=\delta_I^J$. The intersection numbers~$K_{IJK}=D_I \cdot  D_J \cdot  D_K$ were given in~\eqref{int_numb}, which we recall here for convenience
\begin{equation}\label{int_numb_app}
\begin{alignedat}{3}
& \c K_{000} = \base_{\a\b} K^\a K^\b \, , \qquad \quad  && \c K_{00\a} && =  \base_{\a\b} K^\b,\\
& \c K_{0\a\b}  = \base_{\a\b} \, ,   \qquad && \c K_{\a\b\gamma} && = 0\,.
\end{alignedat}
\end{equation}
where~$\base_{\a \b}=D^{\rm b}_\a \cdot  D^{\rm b}_\b$ is the intersection form on the base and the~$K^\a$ appear in the expansion of the canonical class of the base 
\begin{equation}
K=-c_1(B_2)=-\sum_\a K^\a D^{\rm b}_\a=-\sum_\a K_\a C'^\a,
\end{equation}
such that ~$K_\a=c_1(B_2) D^{\rm b}_\a$. 

Following the conventions of ref.~\cite{Klemm2012,Gerhardus:2016iot,bloch2016local,Cota:2017aal}, we choose as basis of branes
\begin{equation}\label{basis-branes}
\c O_\ve = (\c O_{Y_3},\c O_E,\c O_{D_\a},\mathcal{C}^\a,\mathcal{C}^0,\c O_{\rm pt})
\end{equation}
where~$\mathcal{C}^J := \iota_!\c O_{C^J}\big(K^{1/2}_{C^J}\big)$. This basis coincides with the one in~\eqref{tilde_basis}.
The Chern-characters of the 4-branes are~\cite{bloch2016local}
\begin{align}
\ch(\c O_{D_I})& =D_I-\mfrac 12 D_I^2+ \mfrac 16 D_I^3 \\
\intertext{which yields}
\begin{split}
\ch(\c O_E)&=E+\mfrac 12 c_1E+\mfrac 16 c_1^2E\\
\ch(\c O_{D_\a})&=D_\a-\mfrac 12 \base_{\a\a}C^0,
\label{4branes}
\end{split}
\end{align}
while~$\ch(\mathcal{C}^I)=C^I$. 

The Chern character for a general brane~$O_\varepsilon$ can be decomposed as follows
\begin{align}
\begin{split}
\ch_0(\c O_\varepsilon)&=n,\\
\ch_1(\c O_\varepsilon)&=n_E E+F,\\
\ch_2(\c O_\varepsilon)&=EB+n_eC^e,\\
\ch_3(\c O_\varepsilon)&=s,
\end{split}
\end{align}
where~$n,n_E,n_e,s\in \mathbb{Q}$ and can be obtained for our basis of branes by comparing these equations with the above Chern-characters in eqs.~\eqref{4branes} .
Upon performing a Fourier-Mukai transformation, the Chern-character of the transformed brane reads  
\cite{andreas2000fourier,andreas2004fourier}
\begin{align}
\begin{split}
\ch_0(S(\c O_\varepsilon))&=n_E,\\
\ch_1(S(\c O_\varepsilon))&=-nE +B-\mfrac 12 n_E \, c_1,\\
\ch_2(S(\c O_\varepsilon))&=\Big (\mfrac 12 \, n\, c_1-F\Big)^2 E+ \Big (s-\mfrac 12 B c_1 E, +\mfrac{1}{12} \,n_E \, c_1^2 E \Big )C^0,\\
\ch_3(S(\c O_\varepsilon))&=-\mfrac{1}{6}n \, c_1^2E-n_e+\mfrac 12 E \, c_1 F.
\end{split}
\end{align}
where~$c_1=\pi^*c_1(B_2)$. 
Applying to the basis~\eqref{basis-branes}, we find
\begin{align}
\begin{split}
\ch(S(\c O_{Y_3})) & =-\ch(O_{E})+ K_\a \ch(\mathcal{C}^\a), \\
\ch(S(\c O_{E})) &=\ch(O_{Y_3}), \\
\ch(S(\c O_{D_\a}))&=-\base_{\a\b}\ch(\mathcal{C}^\b)+\mfrac 12 \big (\base_{\a\a}+K_\a \big )\ch(O_{\rm pt}) ,\\
\ch(S({\mathcal{C}^0}))&=-\ch(O_{\rm pt}), \\
\ch(S(\c O_{\rm pt}))&=\ch(O_{\mathcal{C}^0}).
\end{split}
\end{align}
This implies that the Fourier-Mukai matrix~$S$ acting on the basis of branes~\eqref{basis-branes} takes the following matrix form
\begin{equation}\label{FM-matrix}
S=\left(
\begin{array}{cccccc}
	0 & -1 &      0       &     K_\a      &                     0                     &               0               \\
	1 & 0  &      0       &       0       &                     0                     &               0               \\
	0 & 0  &      0       & -\base_{\a\b} &                     0                     & \mfrac 12 (\base_{\a\a}+K_\a ) \\
	0 & 0  & \base^{\a\b} &       0       & \mfrac 12 (K^\a-\base^{\a\b}\base_{\b\b}) &               0               \\
	0 & 0  &      0       &       0       &                     0                     &              -1               \\
	0 & 0  &      0       &       0       &                     1                     &               0
\end{array}
\right).
\end{equation}
This matrix leaves invariant the pairing~$\pairing$, i.e.~$S^T\pairing S=\pairing$, where~$\pairing$, introduced in~\eqref{eta-lcs}, takes the form
\begin{equation}
\pairing^T=\left(
\begin{array}{cccccc}
	                 0                  & -2 b_\a K^\a-2 b_0-\frac 16 \K{000} &            -2 b_\a            &       0       & 0 & -1 \\
	2 b_\b K^\b + 2b_0+\frac 16 \K{000} &                  0                  & \frac 12 (\K{00\a}-\K{0\a\a}) &     K^\a      & 1 & 0  \\
	               2b_\b                &    \frac 12 (\K{0\b\b}-\K{00\b})    &               0               & \delta_{\a\b} & 0 & 0  \\
	                 0                  &                -K^\b                &        -\delta_{\a\b}         &       0       & 0 & 0  \\
	                 0                  &                 -1                  &               0               &       0       & 0 & 0  \\
	                 1                  &                  0                  &               0               &       0       & 0 & 0
\end{array}
\right)
\end{equation}
in the basis~\eqref{basis-branes}.

In the main text, we use a basis of branes different than~\eqref{basis-branes}, namely we use the K\"ahler cone basis~\eqref{basis_cO}
\begin{equation}
\c O_\ve^\prime =(\c O_{Y_3},\c O_{D_0},\c O_{D_i},\mathcal{C}^J,\mathcal{C}^0,\c O_{\rm pt})
\end{equation}
containing~$O_{D_0}$ instead of~$O_{E}$. 
Those two basis are related by~\eqref{Kcone}, which in terms of the dual divisors reads
\begin{equation}
E=D_0- K^\a D_\a.
\end{equation}
In matrix notation, this change of basis takes the form 
\begin{equation}
T=\left(
\begin{array}{cccccc}
1 & 0 &   0   & 0 & 0 & 0 \\
0 & 1 & -K^\a & 0 & 0 & 0 \\
0 & 0 &   1   & 0 & 0 & 0 \\
0 & 0 &   0   & 1 & 0 & 0 \\
0 & 0 &   0   & 0 & 1 & 0 \\
0 & 0 &   0   & 0 & 0 & 1
\end{array}
\right).
\end{equation}
The Fourier-Mukai transformation in this basis, given by~$S'=T^{-1}ST$, reads 
\begin{equation}
S'^T=\left(
\begin{array}{cccccc}
	   0     &                 1                  &              0               &                  0                   & 0  & 0 \\
	   -1    &                 0                  &              0               &                  0                   & 0  & 0 \\
	  K^\a   &                 0                  &              0               &             \base^{\a\b}             & 0  & 0 \\
	\K{00\a} &               -K_\a                &        -\base_{\a\b}         &                  0                   & 0  & 0 \\
	   0     &                 0                  &              0               & \frac 12 (K^\a-\base^{\a\b}\K{00\b}) & 0  & 1 \\
	   0     & \frac 12 K^\a (\K{00\a}+\K{0\a\a}) & \frac 12(\K{00\a}+\K{0\a\a}) &                  0                   & -1 & 0
\end{array}
\right)
\end{equation}
where we have displayed the transpose matrix for convenience in the paper. Notice that, in this derivation, we have considered that the coefficient matrix of the branes transforms when going to the small fiber regime, while in the main text we work all the time assuming that the basis transform instead. In practice, this implies that we should work with~$S^T$ instead of~$S$. 

\bibliographystyle{jhep}
\bibliography{emergence}

\end{document}